\documentclass[10pt,conference,review]{IEEEtran}
\IEEEoverridecommandlockouts
\def\BibTeX{{\rm B\kern-.05em{\sc i\kern-.025em b}\kern-.08em
    T\kern-.1667em\lower.7ex\hbox{E}\kern-.125emX}}

\usepackage{algorithmic}
\usepackage{algorithm}
\usepackage{array}
\usepackage{textcomp}
\usepackage{stfloats}
\usepackage[hyphens]{url}

\usepackage{verbatim}
\usepackage{graphicx}
\usepackage{subfigure}
\usepackage{xcolor}

\usepackage{ifthen}

\usepackage[hidelinks]{hyperref}
\newboolean{hidecomments}
\setboolean{hidecomments}{false}
\ifthenelse{\boolean{hidecomments}}
{\newcommand{\nb}[2]{}}
{\newcommand{\nb}[2]{
    \fbox{\bfseries\sffamily\scriptsize#1}
    {\sf\small$\blacktriangleright$ %\RHD\RHD
      {#2} $\blacktriangleleft$}}} %\LHD$}}}%

\usepackage[most]{tcolorbox}
\newcommand{\rqbox}[1]{

\begin{tcolorbox}[tile, size=fbox, boxsep=2mm, boxrule=0pt, top=0pt, bottom=0pt,
borderline west={1mm}{0pt}{gray!50!white}, colback=gray!10!white]
#1
\end{tcolorbox}

}

\usepackage{tikz}

\newcommand*\circledten[1]{\tikz[baseline=(char.base)]{
            \node[shape=circle,fill,inner sep=0.5pt] (char) {\textcolor{white}{#1}};}}

\newcommand{\inlinecircle}[1]{%
    \begin{tikzpicture}[baseline=-0.6ex, scale=0.8]
        \fill[black] (0,0) circle(0.7em);
        \node[white] at (0,0) {\textbf{\normalsize #1}};
    \end{tikzpicture}%
}

\usepackage{balance}

\usepackage{tikz}

\usepackage[fixed]{fontawesome5}

\usepackage{graphicx}
\usepackage{subcaption}
\usepackage{booktabs}
\usepackage{multirow}
\usepackage{array}
\usepackage{listings}
\usepackage{listings, xcolor}

\definecolor{verylightgray}{rgb}{.97,.97,.97}

\lstdefinelanguage{Solidity}{
	keywords=[1]{anonymous, assembly, assert, balance, break, call, callcode, case, catch, class, constant, continue, constructor, contract, debugger, default, delegatecall, delete, do, else, emit, event, experimental, export, external, false, finally, for, function, gas, if, implements, import, in, indexed, instanceof, interface, internal, is, length, library, log0, log1, log2, log3, log4, memory, modifier, new, payable, pragma, private, protected, public, pure, push, require, return, returns, revert, selfdestruct, send, solidity, storage, struct, suicide, super, switch, then, this, throw, transfer, true, try, typeof, using, value, view, while, with, addmod, ecrecover, keccak256, mulmod, ripemd160, sha256, sha3}, % generic keywords including crypto operations
	keywordstyle=[1]\color{blue},
	keywords=[2]{address, bool, byte, bytes, bytes1, bytes2, bytes3, bytes4, bytes5, bytes6, bytes7, bytes8, bytes9, bytes10, bytes11, bytes12, bytes13, bytes14, bytes15, bytes16, bytes17, bytes18, bytes19, bytes20, bytes21, bytes22, bytes23, bytes24, bytes25, bytes26, bytes27, bytes28, bytes29, bytes30, bytes31, bytes32, enum, int, int8, int16, int24, int32, int40, int48, int56, int64, int72, int80, int88, int96, int104, int112, int120, int128, int136, int144, int152, int160, int168, int176, int184, int192, int200, int208, int216, int224, int232, int240, int248, int256, mapping, string, uint, uint8, uint16, uint24, uint32, uint40, uint48, uint56, uint64, uint72, uint80, uint88, uint96, uint104, uint112, uint120, uint128, uint136, uint144, uint152, uint160, uint168, uint176, uint184, uint192, uint200, uint208, uint216, uint224, uint232, uint240, uint248, uint256, var, void, ether, finney, szabo, wei, days, hours, minutes, seconds, weeks, years},	% types; money and time units
	keywordstyle=[2]\color{teal},
	keywords=[3]{block, blockhash, coinbase, difficulty, gaslimit, number, timestamp, msg, data, gas, sender, sig, value, now, tx, gasprice, origin},	% environment variables
	keywordstyle=[3]\color{violet},
	sensitive=true,
	comment=[l]{//},
	morecomment=[s]{/*}{*/},
	stringstyle=\color{red},
	morestring=[b]',
	morestring=[b]"
}

\usepackage{tablefootnote}

\usepackage{marvosym}

\title{Why Is My Transaction Risky? Understanding Smart Contract Semantics and Interactions in the NFT Ecosystem}

\author{
{\rm Yujing Chen}\textsuperscript{1},
{\rm Xuanming Liu}\textsuperscript{1},
{\rm Zhiyuan Wan}\textsuperscript{1}\textsuperscript{*}\textsuperscript{$\dagger$},
{\rm Zuobin Wang}\textsuperscript{1},
{\rm David Lo}\textsuperscript{2},
{\rm Difan Xie}\textsuperscript{3},
{\rm Xiaohu Yang}\textsuperscript{1} \\
\textsuperscript{1}\textit{The State Key Laboratory of Blockchain and Data Security, Zhejiang University} \\
\textsuperscript{2}\textit{School of Computing and Information Systems, Singapore Management University} \\
\textsuperscript{3}\textit{Hangzhou High-Tech Zone (Binjiang) Institute of Blockchain and Data Security} \\
{\tt{\{chenyujing, hinsliu, wanzhiyuan, wangzuobin, yangxh\}@zju.edu.cn}}, \\
{{\tt{davidlo@smu.edu.sg}, \tt{xiedifan@bcds.org.cn}}}
\thanks{* Zhiyuan Wan is the corresponding author.}
\thanks{$\dagger$ Also with Hangzhou High-Tech Zone (Binjiang) Institute of Blockchain and Data Security.}
}

\begin{document}
\maketitle

\begin{abstract}
The NFT ecosystem represents an interconnected, decentralized environment that encompasses the creation, distribution, and trading of Non-Fungible Tokens (NFTs), where key actors, such as marketplaces, sellers, and buyers, utilize smart contracts to facilitate secure, transparent, and trustless transactions.
Scam tokens are deliberately created to mislead users and facilitate
financial exploitation, posing significant risks in the NFT ecosystem.
Prior work has explored the NFT ecosystem from various perspectives, including security challenges, actor behaviors, and risks from scams and wash trading, leaving a gap in understanding the semantics and interactions of smart contracts during transactions, and how the risks associated with scam tokens manifest in relation to the semantics and interactions of contracts.
To bridge this gap, we conducted a large-scale empirical study on smart contract semantics and interactions in the NFT ecosystem, using a curated dataset of nearly 100 million transactions across 20 million blocks on  Ethereum.
We observe a limited semantic diversity among smart contracts in the NFT ecosystem, dominated by proxy, token, and DeFi contracts. 
Marketplace and proxy registry contracts are the most frequently involved in smart contract interactions during transactions, engaging with a broad spectrum of contracts in the ecosystem. 
Token contracts exhibit bytecode-level diversity, whereas scam tokens exhibit bytecode convergence. Certain interaction patterns between smart contracts are common to both risky and non-risky transactions, while others are predominantly associated with risky transactions.
Based on our findings, we provide recommendations to mitigate risks in the blockchain ecosystem, and outline future research directions.

\end{abstract}


\section{Introduction}
A Non-Fungible Token (NFT) is a digital certificate of ownership, immutably recorded on a blockchain like Ethereum. 
While NFTs are commonly associated with digital assets like images and videos, their application has expanded to physical assets, such as postage stamps~\cite{decrypt2021,prnewswire2021}, gold~\cite{cointelegraph2021}, real estate~\cite{blockchainappfactory}, and tangible artwork~\cite{flipkick}, reflecting their growing popularity in diverse markets.
In cryptocurrency, an NFT functions analogously to traditional proof-of-purchase mechanisms, such as invoices or receipts. What distinguishes NFTs is their inherent verifiability and their ability to enable trustless transactions~\cite{wang2021}. 
Verifiability ensures transparent ownership transfer records on the blockchain, enabling clear provenance tracking. Moreover, NFT allows exchanges of digital assets without mutual trust, as both the asset transfer and the cryptocurrency payment are executed atomically within a single, secure transaction on the blockchain.
In 2023, the trading volume of NFTs reached approximately 11.8 billion USD in cryptocurrency~\cite{coingecko2023annual}.

The NFT ecosystem is the interconnected, decentralized environment surrounding the creation, distribution, and trading of NFTs~\cite{das2022understanding}. Key actors, such as marketplaces, sellers, buyers, and content creators, leverage blockchain transactions and decentralized applications (DApps) to facilitate secure, transparent, and trustless exchanges of NFTs. 
Practitioners utilize smart contracts to build DApps, automating relevant processes like NFT creation and transfer in marketplaces~\cite{yang2023definition}. 
Consequently, smart contracts interact with one another to facilitate NFT transactions.
\emph{Scam tokens} are fraudulent or deceptive cryptocurrency assets, implemented as smart contracts, deliberately created to mislead users and facilitate financial exploitation~\cite{wu2024tokenscout}. Scam tokens can typically be classified into three categories: 
1) {\textit{Rugpull} tokens~\cite{2024rugpull,cernera2023token} refer to malicious tokens whose creators deliberately withdraw liquidity or abandon corresponding projects after attracting user investment, leaving holders with worthless assets; 2) \textit{Honeypot} tokens~\cite{torres2019art} lure users into purchasing assets that appear tradable but cannot be resold due to restrictive conditions embedded in token contracts, such as excessive transaction fees, blacklists, or non-standard token logic; 3) \textit{Ponzi} tokens~\cite{2024ponzi} sustain their value by relying on the continuous inflow of funds from new investors to pay returns to earlier ones, with the scheme collapsing when the inflow of new investment slows, resulting in significant losses for the majority of holders.
In 2023, scam tokens caused financial losses of 5.6 billion USD~\cite{business2024losses}, posing significant risks in the blockchain ecosystem.

Previous studies have explored the NFT ecosystem from various perspectives, including security challenges~\cite{das2022understanding}, user activities on prominent NFT marketplaces~\cite{white2022characterizing}, and on-chain behaviors of NFTs~\cite{huang2024unveiling}.
Despite these efforts, a critical gap remains in understanding the semantics of smart contracts in the ecosystem, and how they interact during transactions.
Further research has investigated specific risks in the NFT ecosystem, such as risks from rug pulls~\cite{huang2023miracle}, wash trading~\cite{von2022nft,la2023game}, promotion scams~\cite{roy2024unveiling}, and phishing scams~\cite{yang2024stole}.
However, how the risks associated with scam tokens manifest in relation to the semantics and interactions of smart contracts remain unclear. An in-depth understanding of smart contract semantics and interactions could provide valuable insights into the development practices of smart contracts, as well as inform the design of strategies and tools for detecting and mitigating risks in the blockchain ecosystem.

To address the gaps, we conducted a large-scale empirical study to explore the semantics and interactions of smart contracts in NFT transactions, using a curated dataset comprising nearly 100 million NFT transactions distributed across 20 million blocks on the Ethereum blockchain.
We investigated the following research questions:

\noindent\textbf{RQ1. What are the smart contracts involved in the transactions of the NFT ecosystem?} 

To understand the semantics of smart contracts in the ecosystem, we grouped smart contracts involved in NFT transactions into distinct semantic clusters with respect to their bytecode and source code, and explored the evolution of the number of deployed smart contracts across these semantic clusters over time.
We identified 2,737 semantic clusters of smart contracts in the NFT ecosystem. Nonetheless, the semantic diversity of smart contracts in the ecosystem is limited, as the top 50 largest clusters account for 84.9\% of the total smart contracts. Proxy, token, and DeFi contracts dominate the top 10 largest semantic clusters. Notably, minimal proxy contracts have experienced widespread adoption over an extended period, while financial utility contracts, due to their persistent presence, serve as a foundational layer in the long-term infrastructure of the ecosystem.

\noindent\textbf{RQ2. How do the smart contracts interact during  NFT transactions?}

To capture the interactions among smart contracts during transactions, we measured the frequency and complexity of interactions across the semantic clusters of smart contracts in the NFT ecosystem.
We observed that the semantic clusters with the most frequent interactions during transactions primarily comprise smart contracts related to marketplaces, proxy registry, and proxy.
In the meantime, the transactions tend to involve a limited number of interactions among smart contracts, with a median of four contracts per transaction.
Moreover, the top ten most frequent interaction patterns among smart contracts in transactions exhibit varying levels of complexity, typically associated with specific token operations and business processes.

\noindent\textbf{RQ3. How do scam token risks manifest with respect to the semantics and interactions of smart contracts during NFT transactions?}

To answer RQ3, we conducted a comparison of the bytecode of scam and non-scam tokens, and characterized the interactions of smart contracts in the risky transactions that involve scam tokens.
We found that token contracts demonstrate considerable bytecode-level diversity, with scam tokens exhibiting notable bytecode convergence.
Certain frequently observed interaction patterns between smart contracts are prevalent in both risky and non-risky transactions, while others are predominantly associated with risky transactions, characterized by isolated surges in transaction volumes over distinct short intervals from April 2018 to July 2024.

Based on the findings, we discuss the implications and provide recommendations for mitigating risks in the NFT ecosystem. 
In addition, we outline several research avenues, including real-time monitoring of proxy contracts, and the integration of code- and interaction-level features of smart contracts to enhance fraud detection and transaction risk assessment. 
This paper makes the following contributions:
\begin{itemize}
    \item We present the first large-scale empirical study of smart contract semantics and interactions in the NFT ecosystem on Ethereum, and how the risks associated with scam tokens manifest with respect to the semantics and interactions of smart contracts during transactions.
    \item We curate a dataset that comprises 99,212,864 and 148,411,324 external and internal transactions in the NFT ecosystem, respectively, as well as 225,350 ERC721 contracts for future investigation by others.
    \item We provide practical recommendations for mitigating risks in the NFT ecosystem, and outline avenues of future research.
\end{itemize}

Our replication package is available online: \url{https://doi.org/10.5281/zenodo.15550314}.

\section{Background}\label{sec:background}
\noindent\textbf{EVM-Based Blockchains.} 
The Ethereum Virtual Machine (EVM) serves as the computational backbone of Ethereum and other EVM-compatible blockchains, enabling the execution of smart contracts in a decentralized and trustless manner. Designed as a Turing-complete virtual machine, the EVM allows developers to deploy and execute self-executing code written in programming languages such as Solidity and Vyper, which are compiled into EVM bytecode. The EVM has been widely adopted by numerous blockchain platforms, including Binance Smart Chain, Polygon, Avalanche C-Chain, and Fantom.

EVM-based blockchains operate under an account-based model, distinguishing between \textit{externally owned accounts} (EOAs), which are controlled by users, and \textit{contract accounts} (CAs), which are governed by smart contracts. Transactions on an EVM-based blockchain modify the global state of the blockchain, which is maintained and validated by a distributed network of nodes operating under Proof-of-Work or Proof-of-Stake consensus mechanisms.

\noindent\textbf{Ethereum Tokens.} Ethereum supports the creation and management of fungible and non-fungible assets through standardized token protocols, which are implemented as smart contracts that define and enforce token issuance, transfer, and interaction rules. 
The most widely adopted Ethereum token protocols include ERC20, which facilitates fungible token implementations for DeFi and utility tokens, and ERC721, which enables the creation of unique, non-fungible tokens (NFTs) for digital ownership applications.
Ethereum tokens play a foundational role in tokenized economies, powering DeFi protocols, NFT marketplaces, governance mechanisms, and cross-chain interoperability. 

\noindent\textbf{External and Internal Transactions on Ethereum.} 
Ethereum, as an EVM-based blockchain, facilitates transactions that encompass token transfers, contract deployment, and execution, which can be broadly classified into \emph{external} and \emph{internal} transactions~\cite{wu2021defiranger}. External transactions originate from EOAs, and are uniquely identified by transaction hashes as recorded on chain. EOAs serve as the entry point for the execution of external transactions, enabling the invocation of smart contracts or the transfer of Ether.
In contrast, internal transactions, commonly referred to as the \emph{message calls} in the execution model of Ethereum,  occur during the execution of smart contracts. 
Unlike external transactions, internal transactions are not explicitly recorded on-chain as standalone entities. Instead, reconstructing internal transactions requires the execution traces that include contract invocations (e.g., via {\tt CALL} or {\tt DELEGATECALL} opcode) and \emph{event logs} (e.g., ERC20 or ERC721 {\tt Transfer}  event) which serves to emit structured information that helps track and identify specific contract activities.

\section{Methodology}\label{sec:methodology}
This section outlines the data we collected, and the methodology used to address the research questions, as depicted in Fig.~\ref{fig:methodology}.
\begin{figure*}[htbp]
    \centering
    \includegraphics[width=\textwidth]{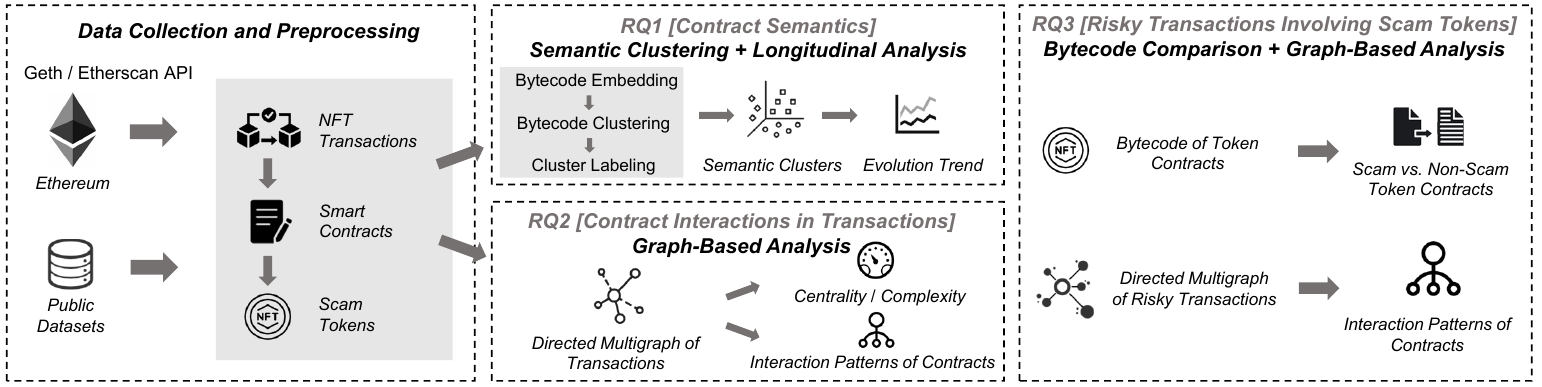}
    \caption{Overview of research methodology.}
    \label{fig:methodology}
\end{figure*}

\subsection{Data Collection and Preprocessing}

\noindent\textbf{NFT Transactions.}
We synchronized Ethereum mainnet blocks up to June 1, 2024, using Geth~\cite{geth}, from block 0 to block 20,000,000. From the event logs recorded in the synchronized blocks, we filtered out ERC721 {\tt Transfer} events and extracted the corresponding external transactions. For each external transaction, we collected the transaction hash, the EOA address that initiated it, and the CA address invoked by the EOA. As a result, we collected 99,212,864 external transactions.
To further collect relevant internal transactions, we located 148,411,324 internal transactions from the XBlock-ETH dataset~\cite{zheng2020xblock} associated with the collected external transactions.
For each internal transaction, we extracted their timestamp, as well as the \texttt{from} and \texttt{to} addresses representing from where and to which account the transaction was sent. Additionally, we extracted the execution order between internal
transactions.

\noindent\textbf{Smart Contracts.} 
Based on the 5,965,248 addresses of CAs involved in our NFT transaction data, we further retrieved their corresponding bytecodes from the XBlock-ETH dataset.
We also collected the source code of the CAs whenever available via the API provided by Etherscan~\cite{etherscan}.
As a result, we collected the source code for 5,108,420 CAs, while the remaining 856,828 were not open-sourced.

\noindent\textbf{Scam Tokens.}
Starting from a public dataset of scam tokens on Ethereum in a recent study~\cite{wu2024tokenscout}, we labeled the 225,350 ERC721 contracts involved in our transaction data.
Specifically, we labeled 3,144 ERC721 contracts as scam tokens, including 2,600 rugpull, 392 honeypot, and 152 Ponzi tokens.

\subsection{Data Analysis}
We seek to answer the following research questions:

\noindent\textbf{RQ1. What are the smart contracts involved in the transactions of the NFT ecosystem?}

To understand the semantics of the smart contracts in the NFT ecosystem, we first captured the \emph{linguistic topics}~\cite{kuhn2007semantic} of smart contracts involved in the NFT-relevant transactions, which reveal the intention of code in smart contracts.
Specifically, we analyzed both bytecode and source code (wherever available) of smart contracts by following three steps: 
1)~\textit{\textbf{Bytecode Embedding}}. 
    We began with the bytecodes of the 5,965,248 smart contracts in our dataset, aiming to capture bytecode embeddings representing smart contracts. Initially, we identified duplicate smart contracts with identical bytecodes, leaving 213,625 unique smart contracts. We then disassembled each of the 213,625 smart contract bytecodes into a sequence of EVM opcodes using Go Ethereum v1.14.7~\cite{geth}. 
    With opcode sequences as input, we further applied the BGE embedding model~\cite{bge} to learning bytecode representation for each smart contract, outputting a 1024-dimensional representation vector. 
    The BGE embedding model has demonstrated its efficacy in generating embeddings for subsequent clustering and classification tasks in recent studies (e.g., ~\cite{zhang2025harnessing,liang-etal-2024-controlled,zheng2025gnncontext}).
2)~\textit{\textbf{Bytecode Clustering}}.
To group smart contract bytecodes, we used the non-parametric clustering algorithm HDBSCAN \cite{campello2013hdbscan}, which has demonstrated its efficacy for code clustering in recent studies (e.g., ~\cite{liang2023needle,ma2025surviving}). HDBSCAN incorporates the idea of hierarchical clustering, thus can automatically select an appropriate density threshold. Specifically, we employed the implementation of the HDBSCAN algorithm from version 0.8.40 of the {\tt hdbscan} library~\cite{hdbscan}.
The Silhouette score~\cite{shahapure2020cluster} is a widely used metric for evaluating the quality of clustering, quantifying how well each data point fits within its assigned cluster by comparing intra-cluster cohesion and inter-cluster separation, with higher values indicating more distinct and well-separated clusters.
To optimize the Silhouette score, we explored multiple combinations of the parameters {\tt min\_cluster\_size} and {\tt min\_samples}, ultimately selecting a value of five for both parameters.
The remaining parameters are set to their default values. 
    As a result, out of the 5,965,248 smart contract bytecodes, we identified 2,737 clusters encompassing 5,478,013 smart contracts, with 525,262 contracts (8.8\%) remaining unclustered. The Silhouette score of 0.827, which is close to the maximum score of 1.0, indicates that each cluster is tightly grouped and well separated from the other cluster.
3)~\textit{\textbf{Cluster Labeling}}.
Through bytecode clustering, we partitioned the smart contracts 
based on their bytecode embeddings into 2,737 clusters. For each cluster $i$, we assigned a label by utilizing the \emph{linguistic topic} derived from the available source code associated with the bytecodes in that cluster. 
Specifically, we first retrieved the source code file(s) associated with all bytecodes within each cluster. 
Next, following prior work~\cite{kuhn2007semantic}, we constructed a term-document matrix $A_i$ by extracting the vocabulary from the source code of each cluster. In the term-document matrix $A_i$, each cluster is represented by a vector of term occurrences, where terms correspond to words appearing in the source code of the cluster. We then calculated the relevance of term $t_j$ to the current cluster $i$ using the formula:
\begin{equation}
    \text{rel}(t_j, i) = \text{sim}(t_j, A_i) - \frac{1}{| \mathcal{A} |} \sum_{A_k \in \mathcal{A}} \text{sim}(t_j, A_k)
\end{equation}
Consequently, we used the top-$n$ list of the most relevant terms from cluster $i$ as its linguistic topic as a reference for labeling. 
We then randomly selected smart contracts with their source code available from each cluster, and scrutinized the source code to infer the underlying semantics of the smart contracts in each cluster.
For the 428 clusters (15.71\%) lacking open-sourced smart contracts, we referred to the information provided for the smart contracts on Etherscan.

Moreover, we investigated the evolution of the numbers of deployed smart contracts across semantic clusters over time. Specifically, we aggregated the smart contracts based on their deployment timestamps in three-month intervals and categorized them according to the corresponding semantic clusters.

\noindent\textbf{RQ2. How do the smart contracts interact during  NFT transactions?}

To capture the interactions among smart contracts during NFT transactions, we introduced a directed multigraph with attributed nodes and edges, denoted as $G = (V,E, \textbf{X}_V,\textbf{X}_E)$. The directed multigraph $G$ consists of (i) a node set $V$ that represents CAs and EOAs on Ethereum associated with NFT transactions, 
(ii) a set of directed edges $E = \{(u, v)\}$ that represents invocations from EOAs to CAs and invocations among CAs during the transactions, 
(iii) a feature vector $x_v \in \textbf{X}_V$ for each node $v \in V$ to represent node attributes, and (iv)  a feature vector $x_e \in \textbf{X}_E$ for each edge $e \in E$ to represent edge attributes.

Given a collection of transactions, we build the directed multigraph as follows.
For each transaction, the involved Ethereum accounts, including both EOAs and CAs, constitute the nodes \( v \in V \), while the invocations between these accounts during the transactions represent the directed edges \( e \in E \) in the multigraph.  
We then leveraged the feature vectors $\mathbf{X}_V$ to capture the semantic characteristics of the nodes. Specifically, for each node representing CA, the feature vector $x_v \in \mathbf{X}_V$ captures its corresponding bytecode embedding, as well as the cluster label, as identified in RQ1. 
Furthermore, we utilized the feature vectors $\mathbf{X}_E$ of edges to capture transaction-level features. Specifically, for each edge $e$ representing an invocation in a transaction, the feature vector $x_e \in \mathbf{X}_E$ captures transaction hash, transaction timestamp, and its position in the invocation sequence during the transaction.

Based on the constructed directed multigraph, we performed an in-depth analysis of the interactions among smart contracts in NFT transactions. 
Specifically, we first aggregated the nodes (smart contracts) in the multigraph based on the semantic clusters identified in RQ1, and then measured the centrality of these semantic clusters in the interactions of smart contracts during transactions.
We further measured the complexity of interactions between smart contracts in the transactions, and employed frequent subgraph mining to extract interaction patterns of smart contracts in the NFT ecosystem.

\noindent\textbf{RQ3. How do scam token risks manifest with respect to the semantics and interactions of smart contracts during NFT transactions?}

To answer RQ3, we started with clustering the 255,350 token contracts in our dataset based on their bytecodes, including 3,144 scam tokens.
Specifically, like RQ1, we first employed the BGE embedding model to derive bytecode representations for each token contract, followed by the application of HDBSCAN to group token contracts with respect to their bytecodes. We then compared the distributions of scam and non-scam tokens across the resulting bytecode clusters of token contracts.

Subsequently, we characterized the interactions of smart contracts during risky transactions involving scam tokens. Specifically, we first sampled subgraphs that represent transactions involving scam tokens from the directed multigraph constructed in RQ2, serving as the risky transactions in the NFT ecosystem.
We then applied frequent subgraph mining to extract interaction patterns of smart contracts in these risky transactions. Additionally, we analyzed the evolution of transaction volumes associated with the extracted interaction patterns of smart contracts in risky transactions.

\section{Results}\label{sec:result}
\subsection{Contract Semantics in the NFT Ecosystem (RQ1)}
Figure~\ref{fig:cluster_dist} illustrates the cumulative distributions of numbers of both (a) total and (b) deduplicated smart contracts per cluster, with clusters ordered by their deduplicated sizes and unclustered smart contracts as an independent group.
Among the 2,737 clusters of smart contracts, the top 10 semantic clusters cover over 30\% of the deduplicated contracts but more than 75\% of the total contracts, indicating that most contracts concentrate on a few dominant semantic patterns while the remainder exhibit semantic diversity in the NFT ecosystem.
Furthermore, as illustrated in Figure~\ref{fig:cluster_dist}a, the cumulative number of smart contracts exhibits two sharp increases upon the inclusion of the clusters ranked 4 and 303, suggesting extensive code duplication within these clusters, which is consistent with the high code clone ratios in smart contracts on Ethereum reported in previous studies~\cite{wang2025clone}.

\begin{figure}[t]
    \centering
    \subfigure[Total]{
        \includegraphics[width=0.45\linewidth]{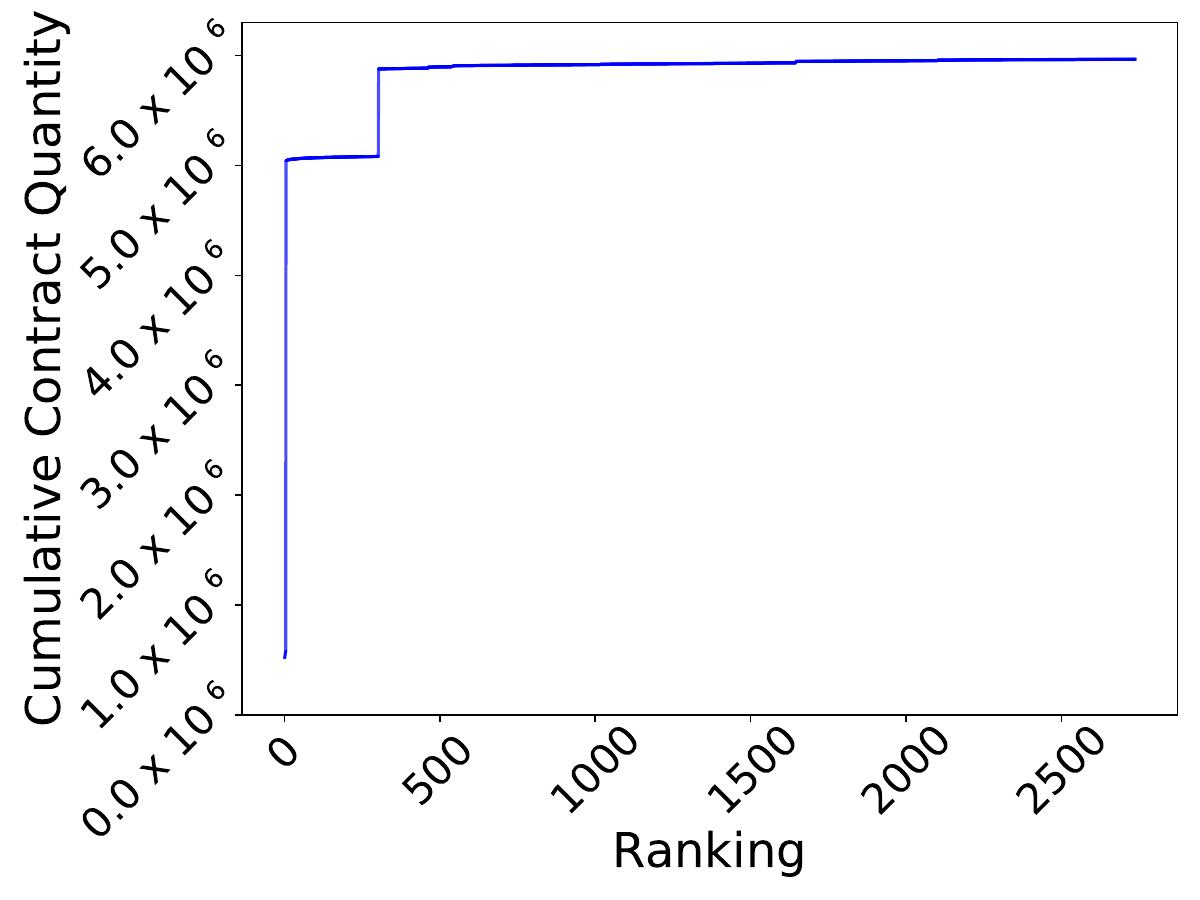}
    }
    \hspace{0.5mm}
    \subfigure[Deduplicated]{
        \includegraphics[width=0.45\linewidth]{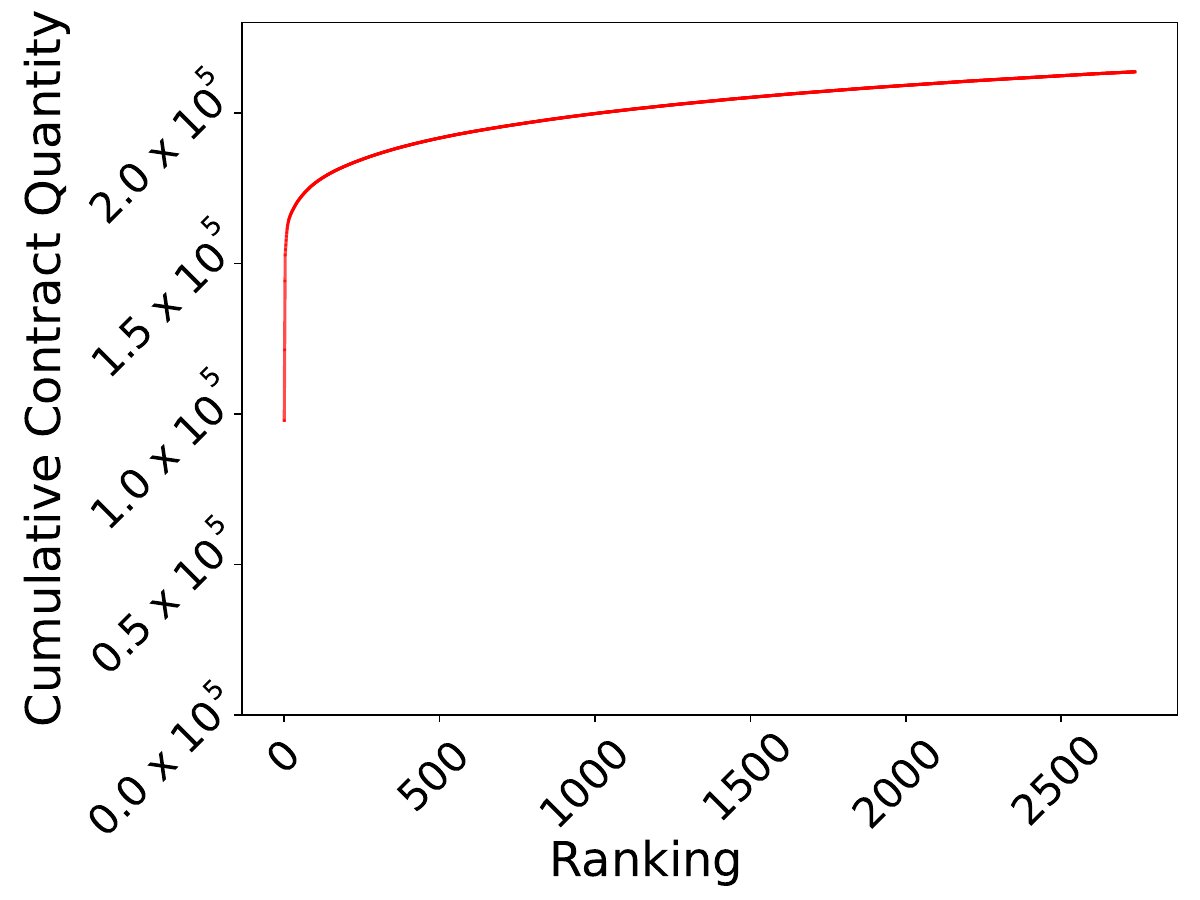}
    }
    \caption{Cumulative number of smart contracts across semantic clusters in the NFT ecosystem.}
    \label{fig:cluster_dist}
\end{figure}

We further conducted an in-depth analysis of the top ten clusters containing the largest number of unique smart contracts in the NFT ecosystem, situated at the leftmost end of the CDF curve. 
Table~\ref{tab:top_ten_clusters} provides a summary of the top ten clusters, which account for 30.4\% of the smart contracts in our dataset. Out of the ten clusters, two do not provide accessible source code (clusters rank 6 and 8), thereby preventing the derivation of corresponding semantic topics from source code (see the Appendix~\cite{replication_package} for detailed linguistic topics).  
Among the top ten clusters of smart contracts, four predominantly comprise proxy contracts, three are associated with DeFi protocols, and the remaining three correspond to NFT assets and protocols. We made the following observations, with example contracts provided in the Appendix~\cite{replication_package}:

\begin{table}[t]
  \centering
  \caption{Top ten semantic clusters with the most smart contracts in the NFT ecosystem.}
  \resizebox{\linewidth}{!}{
    \begin{tabular}{lrc}
    \toprule
    \textbf{Label} & \textbf{\# Deduplicated Contracts} & \textbf{Total (\%)} \\
    \midrule
    \inlinecircle{1} \textsc{Manifold EIP-1967 Proxy A} & 23,425 & 0.39\%  \\
    
    \inlinecircle{2} \textsc{Uniswap v3 Liquidity Pool} & 22,864 & 0.38\%  \\
    
    \inlinecircle{3} \textsc{Manifold EIP-1967 Proxy B}& 8,592 & 0.14\% \\
    
    \inlinecircle{4} \textsc{EIP-1167 Proxy} & 1,867 & 74.82\% \\
    
    \inlinecircle{5} \textsc{Generic EIP-1967 Proxy} & 1,523 & 0.02\% \\
    
    \inlinecircle{6} \textsc{JPEG NFT} & 1,496 & 0.03\% \\
    
    \inlinecircle{7} \textsc{ERC721A NFT} & 1,387 & 0.02\%  \\
    
    \inlinecircle{8} \textsc{Vault} & 1,255 & 0.02\%  \\
    
    \inlinecircle{9} \textsc{ERC721 NFT} & 1,054 & 0.02\%  \\
    
    \circledten{10} \textsc{Financial Utility} & 760 & 0.01\%\\
    
    \bottomrule
    \end{tabular}
  }
  \label{tab:top_ten_clusters}
\end{table}

\noindent\faCubes\textbf{EIP-1967 Proxy Contracts} define a standard for proxy contract storage slots, which specifies how to store the address of the implementation of a contract in a predictable and secure way when using proxy contracts: 

\noindent\faCaretRight\textsc{Manifold EIP-1967 Proxy Contracts} (\inlinecircle{1} and \inlinecircle{3}).
The two clusters both comprise EIP-1967 proxy contracts provided by the \textit{Manifold} protocol~\cite{manifold}. \textit{Manifold} provides toolkits that enable digital creators to mint and manage NFT collections, as well as configurable widgets, APIs, and frameworks that facilitate developers to build NFT experiences~\cite{alchemymanifold}.
The semantic distinction between the two smart contracts primarily lies in their delegation methods.
Specifically, the smart contract from Cluster \inlinecircle{1} uses \texttt{Address.functionDelegateCall} with a security check that verifies the target contract address before executing the delegate call. 
In contrast, the smart contract from Cluster \inlinecircle{3} uses \texttt{delegatecall}, which bypasses such a security check.

\noindent\faCaretRight\textsc{Generic EIP-1967 Proxy Contracts (\inlinecircle{5}).} 
The contracts in Cluster 5 also adhere to the EIP-1967 standard. Meanwhile, they offer greater flexibility by allowing the logic contract address and initialization parameters to be specified as constructor arguments during deployment. 

\noindent\faCubes\textbf{\textsc{EIP-1167 Proxy Contracts (\inlinecircle{4}).}} 
EIP-1167 defines a minimal proxy contract, which provides a standardized way to deploy lightweight contracts that delegate their execution to another contract, often referred to as the \emph{master} contract~\cite{murray2018eip}. The EIP-1167 proxy contracts, with minimal bytecode, contain the logic to delegate calls to existing contracts, thus forwarding any transactions or function calls to the corresponding master contracts.
The cluster consists of 4,463,176 contracts, with 1,867 unique bytecode, indicating extensive code cloning related to EIP-1967 proxy contracts in the NFT ecosystem.

\noindent\faCubes\textbf{NFT Contracts} define and manage NFTs, representing ownership of items such as art, collectibles, music, or in-game items:

\noindent\faCaretRight \textsc{JPEG NFT Contracts (\inlinecircle{6}).}
JPEG mining~\cite{jpegmining} refers to a type of NFT creation in which actual image data is directly uploaded to the blockchain, contrasting with conventional NFTs that typically store only metadata on-chain. During the minting process, miners upload the image data and receive an NFT in return. Contracts in this cluster are generated as part of the JPEG mining process and include information related to the image data of the NFT.

\noindent\faCaretRight \textsc{ERC721 NFT Contracts (\inlinecircle{9}).} 
The cluster comprises NFT contracts built upon the implementation of the ERC721 standard by OpenZeppelin~\cite{erc721oppenzeppelin}, which provides essential functionalities such as token minting and transfer, as well as extended features, including metadata support and approval mechanisms.
ERC721 is the most widely used standard for representing NFTs on the Ethereum blockchain. Each ERC721 token is inherently unique, capable of carrying distinct metadata, ownership, and provenance.

\noindent\faCaretRight \textsc{ERC721A NFT Contracts (\inlinecircle{7}).} 
As the NFT ecosystem continues to expand, an increasing number of projects demand scalable mechanisms for large-scale token issuance. However, the standard ERC721 implementation incurs substantial gas costs when minting tokens in batches. To address this limitation, the ERC721A standard~\cite{erc721a} was introduced as a gas-efficient alternative, specifically designed to optimize batch minting while remaining interface-compatible with ERC721. 
The cluster consists of NFT contracts adopting ERC721A, which implement efficient batch minting along with additional optimizations. 

\noindent\faCubes\textbf{DeFi Contracts} enable decentralized financial services without reliance on traditional intermediaries, extending the utility and financialization of NFTs:

\noindent\faCaretRight\textsc{Uniswap v3 Liquidity Pool Contracts (\inlinecircle{2}).}
Uniswap v3~\cite{uniswapv3} introduces an interaction model with NFTs through the concept of concentrated liquidity, wherein each liquidity position is represented as an NFT. Unlike previous versions of Uniswap, where liquidity providers (LP) received fungible LP tokens, Uniswap v3 assigns an ERC721-compliant NFT to each liquidity position, encoding the owner's specific parameters such as token pair, fee tier, and price range. This design enables more granular control over capital allocation and turns each position into a composable, tradable asset on-chain. 
As a result, these position NFTs can be transferred, collateralized in DeFi protocols, or integrated into broader NFT marketplaces and financial infrastructure. 

\noindent\faCaretRight\textsc{Vault Contracts (\inlinecircle{8}).} ERC721 vault contracts, instantiated via the {\tt mint} function of a {\tt VaultFactory} contract, are designed to custody individual NFTs and issue ERC20 tokens representing fractional ownership. These vaults enable decentralized management and trading of fractionalized NFT assets.

\noindent\faCaretRight\textsc{Financial Utility Contracts for NFTs (\inlinecircle{10}).}
Financial smart contracts involve NFT-backed lending, collateral management, and other financial functionalities. Contracts in this cluster implement mechanisms for interest and debt calculation, loan issuance and repayment, as well as auction-based liquidations and collateral redistribution.  

\begin{figure}[t]
    \centering
        \includegraphics[width=\linewidth]{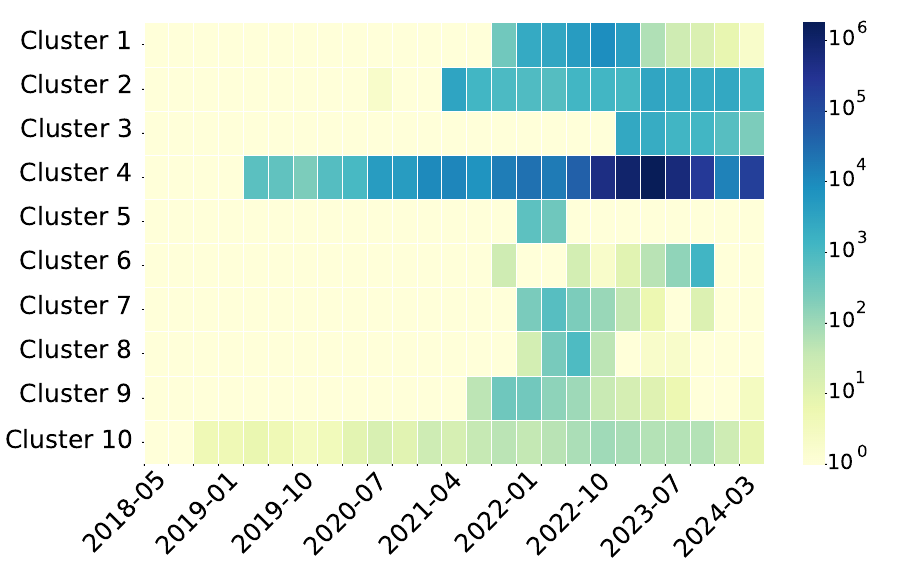}
    \caption{Evolution in the numbers of deployed smart contracts across the top ten semantic clusters.}

    \label{fig:sc_evolution}
\end{figure}
Figure~\ref{fig:sc_evolution} illustrates the quarterly deployment dynamics of the top ten contract clusters in the NFT ecosystem. 
Cluster \inlinecircle{4}, primarily consisting of minimal proxy contracts, dominates in both longevity and volume, reflecting the widespread adoption of cost-efficient, scalable deployment patterns in the NFT ecosystem. 
Cluster \inlinecircle{2}, representing Uniswap V3 Liquidity Pool contracts, reached 13,806 deployments between January 2023 and June 2024, whereas standard NFT contracts such as ERC721A and ERC721 (Clusters \inlinecircle{7} and \inlinecircle{9}) appeared only 106 times, reflecting a broader trend of increasing integration between NFTs and DeFi that enables novel use cases.
In contrast, Cluster \inlinecircle{10}, encompassing financial utilities for NFTs such as lending and auction contracts, maintains a relatively low but stable deployment rate, indicating its role as a long-term infrastructure layer rather than a trend-driven component.

We do not observe any deployments of the ten semantic clusters of smart contracts prior to April 2018.

\rqbox{
\textbf{Finding 1:} 
The semantic diversity of smart contracts in the NFT ecosystem is limited, with the top 50 largest clusters comprising 80.2\% and 84.9\% of the deduplicated and total smart contracts, respectively. Proxy, token, and DeFi contracts dominate the top 10 largest semantic clusters. Notably, minimal proxy contracts have seen widespread adoption over a prolonged period, while financial utility contracts, with their sustained presence, function as a crucial layer in the long-term infrastructure of the ecosystem.
}

\subsection{Contract Interactions in NFT Transactions (RQ2)} \label{sec:RQ2-result}

In this section, we present the results of our analysis of the interactions between smart contracts in NFT transactions, based on the constructed directed multigraph.

\subsubsection{Centrality of Smart Contracts in Interactions}

Table~\ref{tab:top_five_degree_clusters} summarizes the top five semantic clusters of smart contracts in the NFT ecosystem with the highest degree in the directed graph, including two clusters associated with NFT marketplaces, two related to proxy registries, and one dedicated to proxies (see the Appendix~\cite{replication_package} for detailed linguistic topics). 
The two clusters associated with NFT marketplaces server distinct functions: 
Smart contracts in \textsc{Marketplace A} are primarily concerned with high-level operations related to NFT marketplace activities, such as managing sales, pricing strategies, and user roles. In contrast, \textsc{Marketplace B} focuses on the technical infrastructure necessary for processing offers and executing orders, including data handling, execution flows, and error management mechanisms.
In the case of the proxy registry-related clusters, which function as centralized registries for proxy contracts, the smart contracts in \textsc{Proxy Registry A} handle sophisticated state management and execution of proxy operations based on specific conditions. Meanwhile, \textsc{Proxy Registry B} focuses on the core functionality of proxies, including delegated calls and the management of ownership and access control for proxy contracts.
In addition, the smart contracts in the \textsc{Proxy} cluster are primarily concerned with the technical aspects of implementing proxy patterns, focusing on the upgradeability and versioning of smart contracts.

\begin{table}[t]
  \centering
  \caption{Top five semantic clusters of smart contracts with the highest degree of interaction occurring during transactions.}
  \resizebox{0.9\linewidth}{!}{
    \begin{tabular}{lcc}
    \toprule
    \textbf{Label} & \textbf{Degree} & \textbf{\# Contracts} \\
    \midrule
     \textsc{Marketplace A} & 66,092,098 & 29 \\
    
     \textsc{Marketplace B} & 56,193,686 & 7 \\
    
     \textsc{Proxy Registry A} & 48,231,893 & 19 \\
    
     \textsc{Proxy} & 47,640,194 & 795,872 \\
    
     \textsc{Proxy Registry B} & 45,361,729 & 5\\
    
    \bottomrule
    \end{tabular}
  }
  \label{tab:top_five_degree_clusters}
\end{table}

We also observed that the smart contracts that remain unassigned to any semantic clusters have a cumulative degree of 306,999,918 in the directed multigraph.
The top three unclustered contracts by degree are \textit{OpenSea: Conduit}, \textit{ENS: Old ETH Registrar Controller}, and \textit{ENS: Registry with Fallback}. The high degrees of these contracts indicate their frequent interactions with other smart contracts in the NFT ecosystem, suggesting the active and prominent role of OpenSea and Ethereum Name Service (ENS) in the NFT ecosystem.

\rqbox{
\textbf{Finding 2:} 
The semantic clusters with the highest degrees of interaction during transactions in the NFT ecosystem are primarily composed of smart contracts related to NFT marketplaces, proxy registries, and proxies. Furthermore, unclustered contracts associated with OpenSea and Ethereum Name Service also demonstrate frequent interactions during transactions.
}

\subsubsection{Interaction Statistics and Patterns}

According to the statistics derived from the directed multigraph, the median number of interacting smart contracts per transaction is 3 (min: 1, mean: 5.1, max: 603), while the median number of interactions between these contracts is 4 (min: 1, mean: 7.5, max: 12,903).
On one hand, the relatively low median values indicate that, on average, transactions in the NFT ecosystem typically involve limited interactions between smart contracts. 
On the other hand, the wide ranges in both the number of contracts involved (from 1 to 603) and the frequency of interactions (ranging from 1 to 12,903) suggests that a subset of transactions might involve more complex and intricate interactions between contracts.

Moreover, we identified 771,839 distinct interaction patterns between smart contracts in NFT transactions, with the ten most frequently observed patterns depicted in Figure~\ref{fig:transaction_pattern}a. We made the following observations: 

\begin{figure*}[htbp]
    \centering
    \subfigure[Throughout all NFT transactions]{\includegraphics[width=0.485\linewidth]{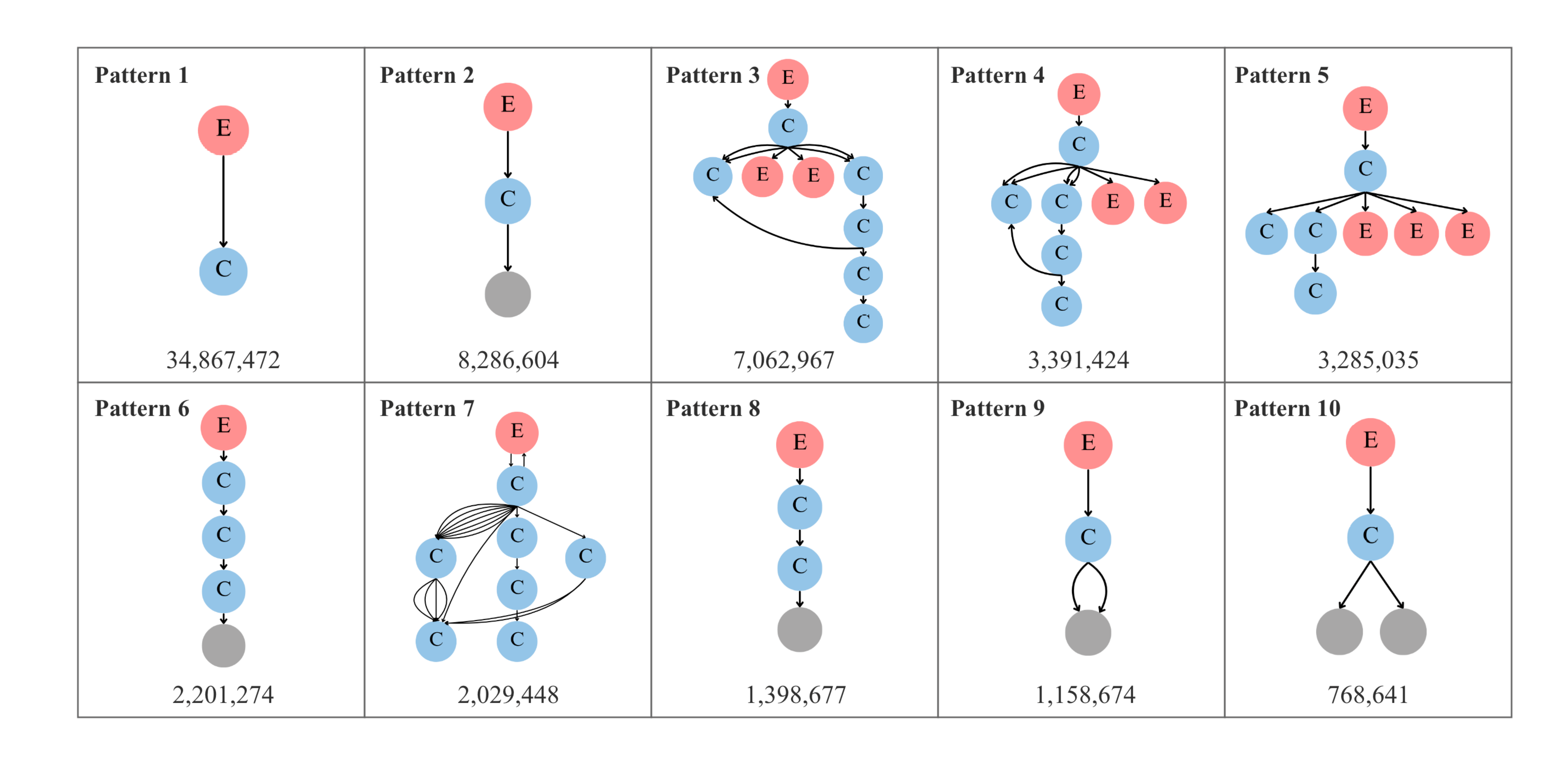}}
    \subfigure[In transactions involving scam tokens (the corresponding \emph{risky transaction ratios} are indicated in parentheses)]{\includegraphics[width=0.485\linewidth]{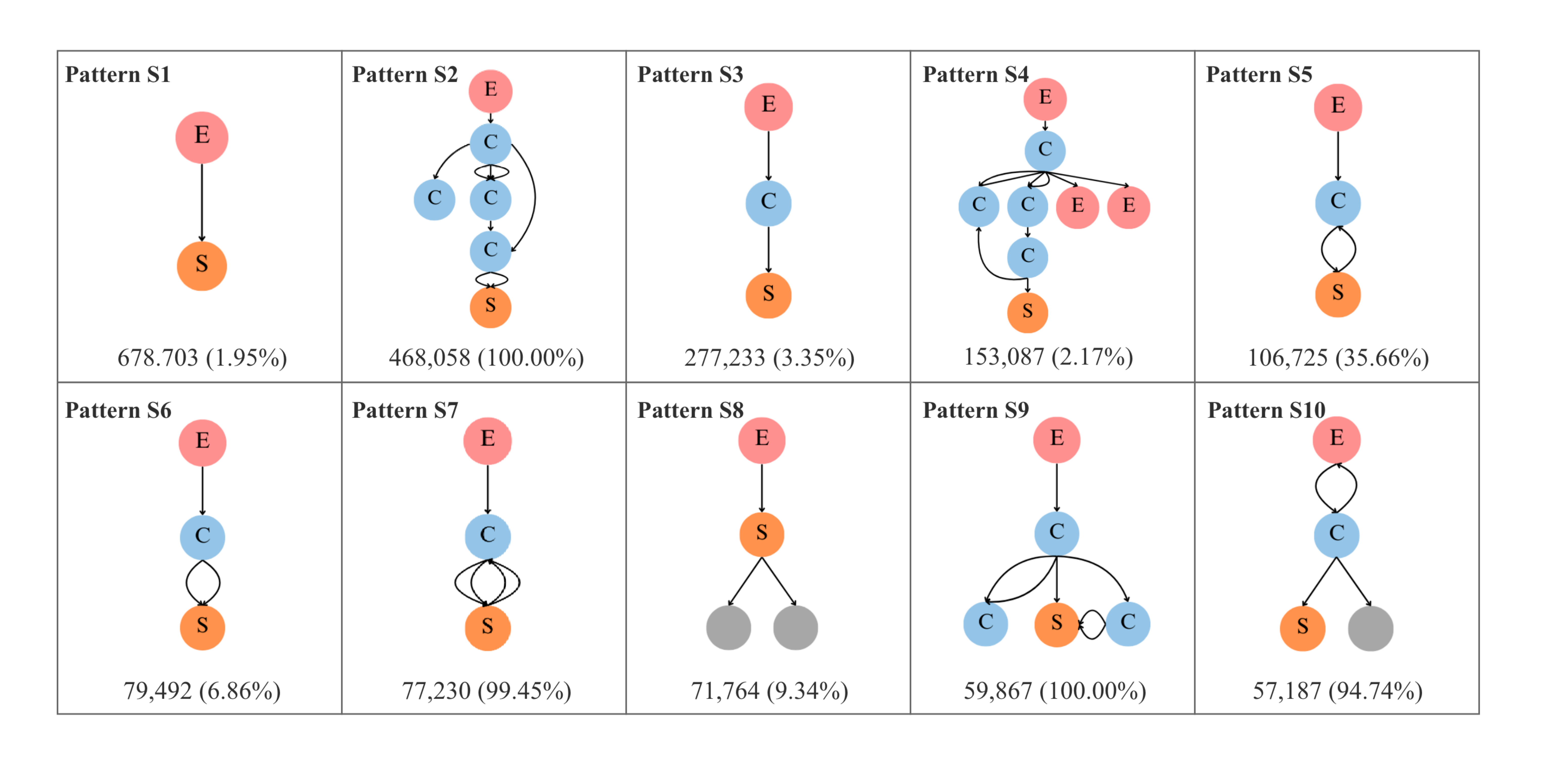}}
     \caption{Top ten most frequent interaction patterns between smart contracts in NFT transactions, where EOAs, CAs, and scam tokens are denoted as E, C, and S, respectively. Grey nodes represent either CAs or EOAs.} 
    \label{fig:transaction_pattern}
\end{figure*}

\noindent\faHandPointRight\textbf{Pattern 1: Minimal Interactions  (35.1\%)} Pattern 1 demonstrates the direct interaction between EOAs and NFT contracts, occurring in over 34 million transactions (35.1\%) in our dataset. Notably, 59.6\% (1357 out of 2276) of the identified semantic clusters of smart contracts in RQ1 are involved in these interactions.

\noindent\faHandPointRight\textbf{Pattern 2 and 8: Chained Interactions (8.4\% and 1.4\%)} 
Pattern 2 and 8 are observed in three types of transactions: 1) an EOA initiates the first NFT smart contract that implements certain extended functionalities, which subsequently calls additional smart contract(s) from distinct semantic clusters; 2) an EOA interacts with a proxy initially, which then delegates the call to an NFT contract, and in some cases, triggers further calls to other contracts; 3) an EOA initially interacts with smart contracts that handle the logic for NFT management, such as NFT auction and NFT staking, ultimately triggering NFT transfers.

\noindent\faHandPointRight\textbf{Pattern 3, 4 and 5: Interactions via Marketplace (7.1\%, 3.4\% and 3.3\%)} 
As the leading NFT marketplace, \emph{OpenSea} has launched two representative protocols for decentralized trading of NFTs,   \textit{Wyvern}~\cite{wyvern} in 2017 and  \textit{Seaport}~\cite{seaport} in 2022.
Patterns 3 and 4 depict frequent interactions between smart contracts in NFT transactions utilizing the \textit{Wyvern} protocol, while Pattern 5 illustrates interactions in transactions involving the \textit{Seaport} protocol.
A comparison of these patterns suggests that transactions utilizing the \textit{Seaport} protocol tend to be more streamlined, characterized by shorter call chains in smart contracts, and exhibit greater flexibility, particularly with multi-asset transactions that involve multiple users, as compared to the \textit{Wyvern} protocol.

\noindent\faHandPointRight\textbf{Pattern 6: \emph{Gnosis Safe} Involved Interactions (2.2\%)} 
\emph{Gnosis Safe} is a platform that offers a Smart Account solution for Ethereum, which provides multi-signature wallets that require multiple parties to approve transactions, thereby enhancing security for users to manage digital assets~\cite{safe}.
Pattern 6 frequently occurs in the transactions involving \emph{Gnosis Safe} where the first interacted contract functions as a proxy, and the second contract, which serves as the proxy's logic contract, primarily consists of NFT contracts with diverse extend features.
The third invoked contract provides methods for the creation and interaction with proxy contracts within the \textit{Gnosis Safe} ecosystem, while the fourth contract implements the core functionality of Gnosis Safe.
In these transactions, users typically purchase NFTs and transfer payment in Ether through a proxy, while the NFT contract subsequently sends a portion of the payment received from the users to the \textit{Gnosis} proxy of a specific platform to cover the fee.

\noindent\faHandPointRight\textbf{Pattern 7: Interactions via ENS (2.1\%)} 
Pattern 7 is frequently observed in transactions involving the Ethereum Name Service (ENS). A typical example of this pattern involves an EOA initiating a transaction by invoking the \textit{ETH Registrar Controller} contract, which subsequently interacts with several ENS-related smart contracts, including: 1) the \textit{Base Registrar Implementation} contract, which facilitates the registration of Ethereum-based domain names; 2) the \textit{Registry with Fallback} contract, representing the ENS registry with an integrated fallback mechanism; and 3) the \textit{Public Resolver} contract, which resolves ENS domains to various associated resources.

\noindent\faHandPointRight\textbf{Pattern 9: Repeated Interactions (1.2\%)} 
Pattern 9 is frequently observed in interactions involving \emph{unclustered} smart contracts in NFT transactions. 
As Pattern 9 illustrates, a smart contract repeatedly invokes another contract in a transaction, typically for purposes such as performing NFT batch operations, confirming and validating states before and after NFT transfers, or setting the state of NFTs.

\noindent\faHandPointRight\textbf{Pattern 10: Branching Interactions (0.8\%)} 
Pattern 10 is frequently observed in transactions involving ether transfers to EOAs, such as those related to the purchase of NFTs, payment to the buyer, or payment of platform fees. For example, in a transaction~\cite{tx_p10} associated with an NFT auction, the auction contract first transfers the Ether received from the buyer to the seller, and then calls the NFT contract to facilitate the transfer of the token to the buyer.

\rqbox{
\textbf{Finding 3:} 
NFT transactions typically involve a limited number of interactions between smart contracts, with a median of four contracts per transaction.
Among the 771,839 distinct interaction patterns identified between smart contracts in NFT transactions, the top ten most frequent patterns exhibit varying levels of complexity and are associated with specific NFT operations and business processes.
}

\subsection{Risky Transactions Involving Scam Tokens (RQ3)}

This section presents the results of our comparison between the bytecodes of scam and non-scam tokens in the NFT ecosystem, along with the interaction patterns between smart contracts in transactions involving scam tokens.

\subsubsection{Scam vs. Non-Scam Tokens}

We identified 1,350 distinct clusters from the bytecodes of 255,350 ERC721 smart contracts (i.e., scam and non-scam tokens) in the NFT ecosystem.
The bytecode clusters of these token contracts vary significantly in size, ranging from 5 to 97,319.
Scam tokens span 45 out of the 1,350 bytecode clusters, with varying ratios across the clusters. 
Table~\ref{tab:propotion_of_scam_in_cluster} shows the distribution of the 45 clusters with scam tokens across various intervals of scam token ratios.
Specifically, among the 45 clusters, 20 clusters consist entirely of scam tokens (100\%), while an additional 18 clusters have a high scam token ratio between 60\% and 100\%. Together, these two categories of clusters account for 2,326 scam tokens, and represent the majority of scam tokens in our dataset.
In contrast, two clusters exhibit a moderate scam token ratio between 40\% and 60\%, comprising 93 scam tokens. Meanwhile, five clusters fall within the [0\%, 40\%) range, with 714 out of 725 scam tokens in this group being unclustered. 
This distribution suggests that scam tokens are not randomly dispersed but tend to concentrate within a limited number of semantically similar clusters, reflecting strong code-level homogeneity among scam tokens.

We further analyzed the available source code of representative bytecode clusters across different scam token ratio intervals, and derived the following observations:

\noindent\faCaretRight
\textbf{Clusters with 100\% Scam Tokens.} 
We selected a representative cluster containing 51 scam tokens, all of which were labeled as rugpull tokens.
As exemplified by the {leftoverz}~\cite{leftoverz} contract, the scam tokens in this cluster typically extend the ERC721 standard by incorporating two additional features: (i) a \emph{whitelist} mechanism that facilitates NFT airdrops to a designated group of users, and (ii) a \emph{payment splitter} component that allocates and distributes payments among multiple recipients according to predefined shares.

\noindent\faCaretRight\textbf{Clusters with [60\%,100\%) Scam Tokens.}
We selected a representative cluster containing 1,426 token contracts, with 96\% of which being labeled as scam tokens.
As exemplified by the \emph{EtherElephants}~\cite{CryptoToon} contract, the token contracts in this cluster tend to be a direct fork of the \texttt{ERC721Full} contract, which is an \textit{outdated} reference implementation of the ERC721 standard provided by OpenZeppelin~\cite{erc721oppenzeppelin}.

\noindent\faCaretRight\textbf{Clusters with [0,60\%) Scam Tokens.}
Among the seven clusters with scam token ratios between 0\% and 60\%, the source code of the tokens exhibits considerable diversity in the ways the ERC721 standard is extended.
The semantic variations in the included tokens suggest a deliberate effort by scammers to explore diverse contract-level mechanisms for executing fraudulent schemes.

\begin{table}[t]
  \centering
  \caption{Statistics of token clusters across ranges of scam token ratios.} 
  \resizebox{\linewidth}{!}{
    \begin{tabular}{>{\centering\arraybackslash}p{6em} >{\centering\arraybackslash}p{6em} >{\centering\arraybackslash}p{10em} >{\centering\arraybackslash}p{8em}}
    \toprule
\textbf{Scam Token Ratio (\%)} & \textbf{\# Clusters} & \textbf{\# Scam Tokens} & \textbf{\# Total Tokens} \\
    \midrule
    $100\%$ & 20 & 321 & 321 \\
    $[60\%,100\%)$ & 18 & 2,005 & 2,143 \\
    $[40\%,60\%)$ & 2 & 93 & 197 \\
    $[0,40\%)$ & 5 & 725 (714 is unclustered) & 77,813 \\
    \bottomrule
    \end{tabular}
  }
  \label{tab:propotion_of_scam_in_cluster}
\end{table}

\rqbox{
\textbf{Finding 4:} 
Token contracts in the NFT ecosystem exhibit significant bytecode-level diversity, distributed across 1,350 distinct clusters, with scam tokens predominantly concentrated in 45 (3.3\%) of these clusters. Among the 45 clusters, 38 demonstrate scam token ratios exceeding 60\%, indicating a high concentration of fraudulent activity in specific bytecode patterns.
}

\subsubsection{Interaction Patterns in Risky Transactions}
We identified 22,306 interaction patterns between smart contracts from the 3,089,245 risky transactions in the NFT ecosystem, with the ten most frequently observed patterns depicted in Figure~\ref{fig:transaction_pattern}b. Through analysis of these patterns, we made the following observations:

\noindent\faHandPointRight\textbf{Patterns S2, S7, S9 and S10 are predominantly associated with risky transactions, with each pattern exhibiting a risky transaction ratio exceeding 90\%.}
Patterns S2, S7, S9 and S10 exhibit high risky transaction ratios of 100\%, 99.45\%, 100\%, and 94.74\%, respectively.
The predominance of these patterns in risky transactions suggests that certain interactions among smart contracts are highly indicative of malicious behavior in the NFT ecosystem.
We also noticed that these patterns are grounded in specific contract-level workflows, including minting, structured transfers, and auction-based bidding mechanisms.
For example, Pattern S2 reflects user interactions during the minting process of the \emph{Codex Record} NFT collection~\cite{codex}.
Pattern S7 is derived from a characteristic transfer behavior observed in the \emph{LucidSight} collection~\cite{MLBNFT}.
Patterns S9 and S10 are associated with the auction mechanisms implemented in the \emph{Marble}~\cite{MarbleNFT} and \emph{CryptoFlower}~\cite{CryptoFlower} collections, respectively.
Notably, Pattern S10 captures a bidding workflow: an EOA places a bid by transferring a fixed amount of ETH and subsequently receives a refund for any surplus once the auction concludes.

\noindent\faHandPointRight\textbf{Patterns S1, S3, S4, S6 and S8 are shared across both risky and non-risky transactions, with each pattern exhibiting a risky transaction ratio below 10\%.}
Patterns S1, S3, S4, S6 and S8 observed in risky transactions correspond to Patterns 1, 2, 4, 9, and 10 identified in RQ2, with each pattern exhibiting a relatively low risky transaction ratio of 1.95\%, 3.35\%, 2.17\%, 6.86\%, and 9.34\%, respectively.
The relatively low risky ratios associated with these five interaction patterns reflect their dual-use nature, as they are adopted by both legitimate users and malicious actors in their transactions in the NFT ecosystem. 
The shared usage of such interaction patterns across transactions may limit the discriminative utility of these patterns in distinguishing between risky and non-risky transactions in the NFT ecosystem. It may require the incorporation of contextual and semantic features when designing effective detection mechanisms for risk transactions exhibiting such patterns.

\begin{figure}[t]
    \centering
    \includegraphics[width=\linewidth]{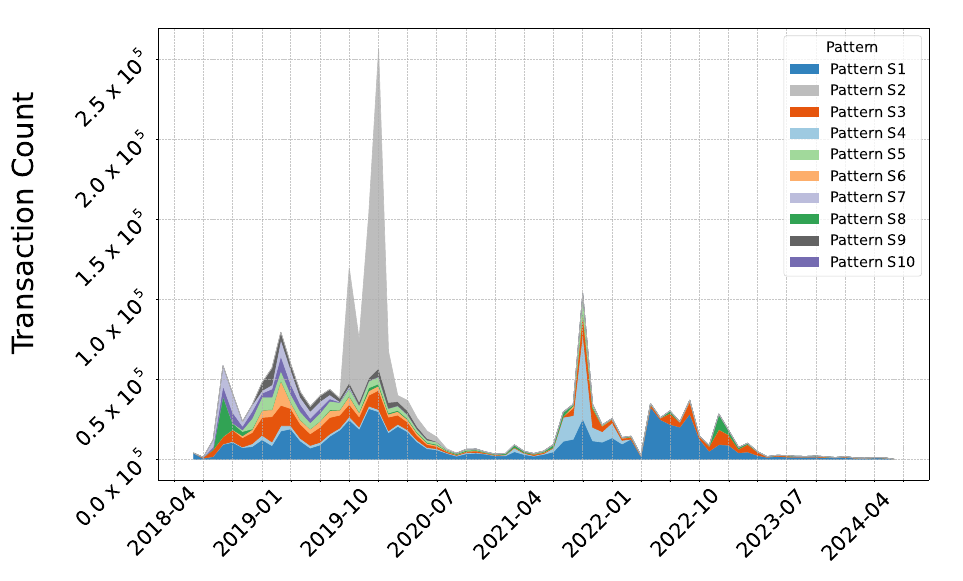}
    \caption{Evolution in transaction volumes for the ten most frequent interaction patterns of scam tokens.}
    \label{fig:transaction_pattern_temporal_stack}
\end{figure}

Figure~\ref{fig:transaction_pattern_temporal_stack} presents the evolution of risky transaction volumes associated with the above ten most frequently observed interaction patterns, spanning the period from April 2018 to July 2024. Among these patterns, Pattern S1 demonstrates sustained activity throughout the entire timeframe, suggesting that the structural generality and flexibility of the interaction pattern have contributed to its continued use in scam campaigns.
In contrast, Pattern S2 exhibits a sharp and isolated surge in transaction volume between late 2019 and mid-2020, peaking at over 200,000 transactions. This suggests a targeted exploitation of specific contract-level mechanisms, likely related to the minting logic of the rugpull-like \textsf{Codex Record} collection.
In addition, Patterns S7 and S10 exhibit modest yet distinct spikes during earlier periods in the timeframe, suggesting their involvement in localized or short-lived scam activities.
Notably, the transaction volumes associated with all patterns demonstrate a consistent decline after early 2022, which may be attributed to the emergence of more effective detection mechanisms, shifts in attacker strategies beyond scam tokens, or a reduced economic incentive for conducting large-scale exploitation.

\begin{figure}[t]
    \centering
    \subfigure[Rugpull]{\includegraphics[width=0.23\linewidth]{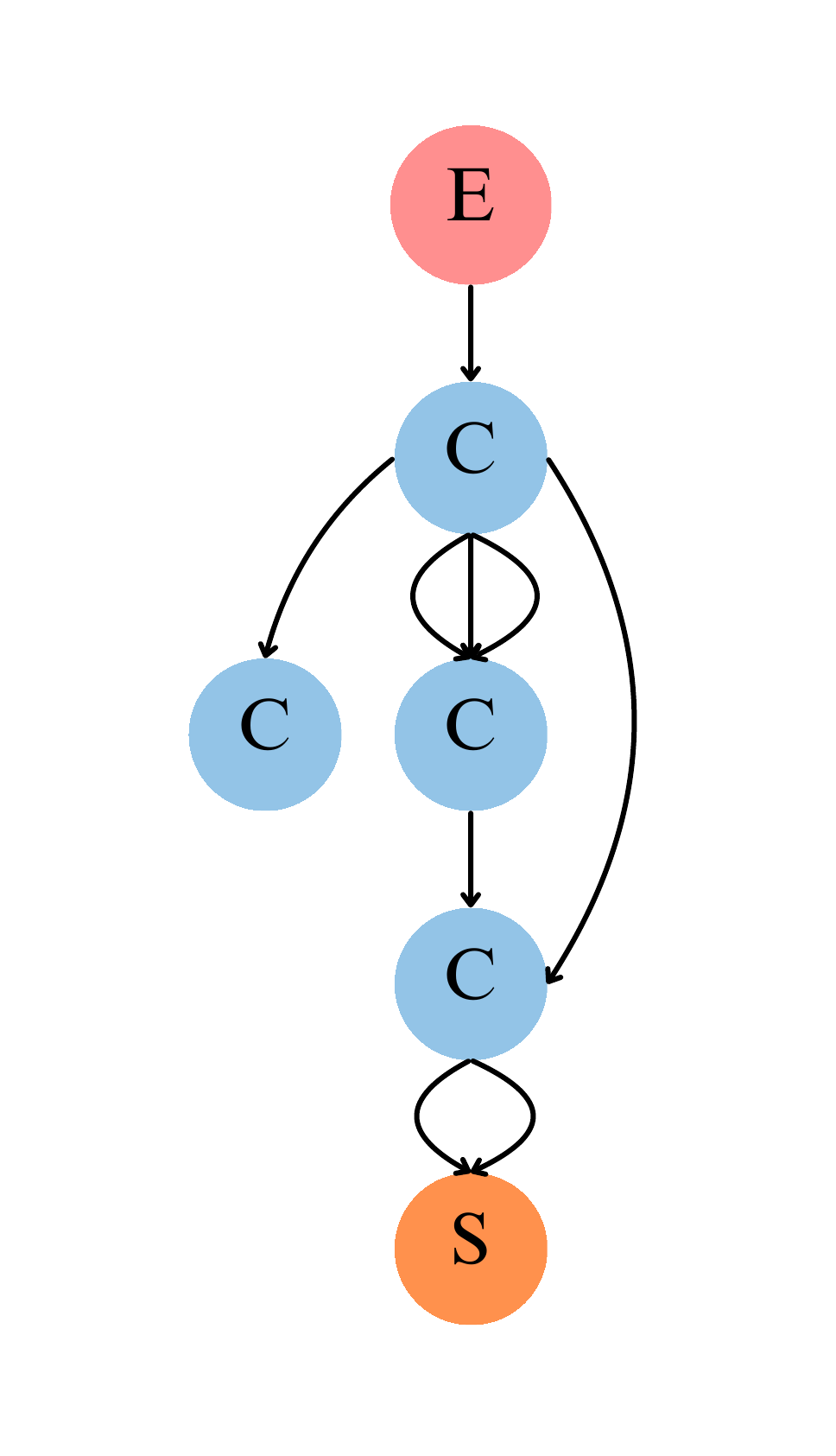}}
    \subfigure[Honeypot]{\includegraphics[width=0.32\linewidth]{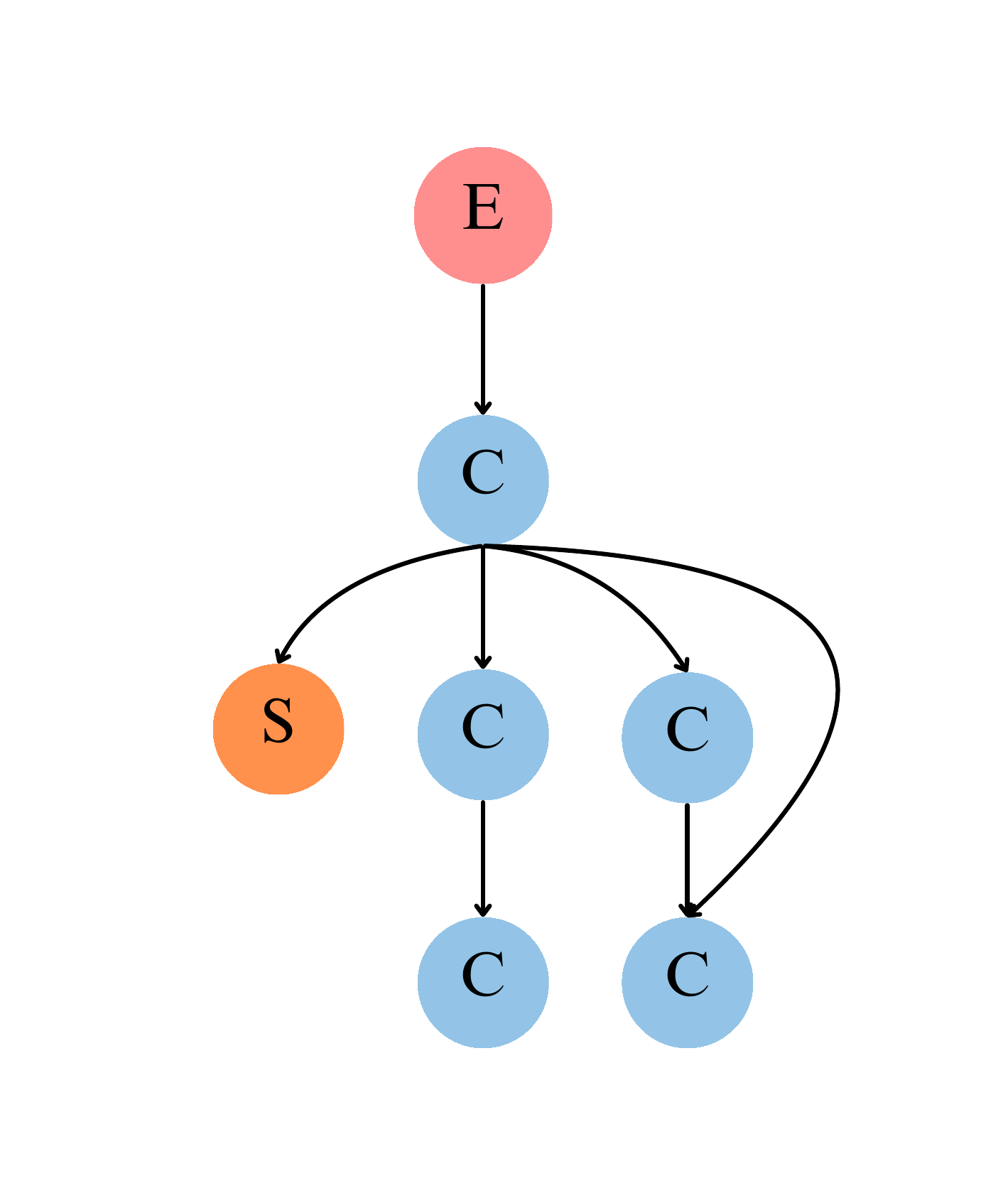}}
    \subfigure[Ponzi]{\includegraphics[width=0.38\linewidth]{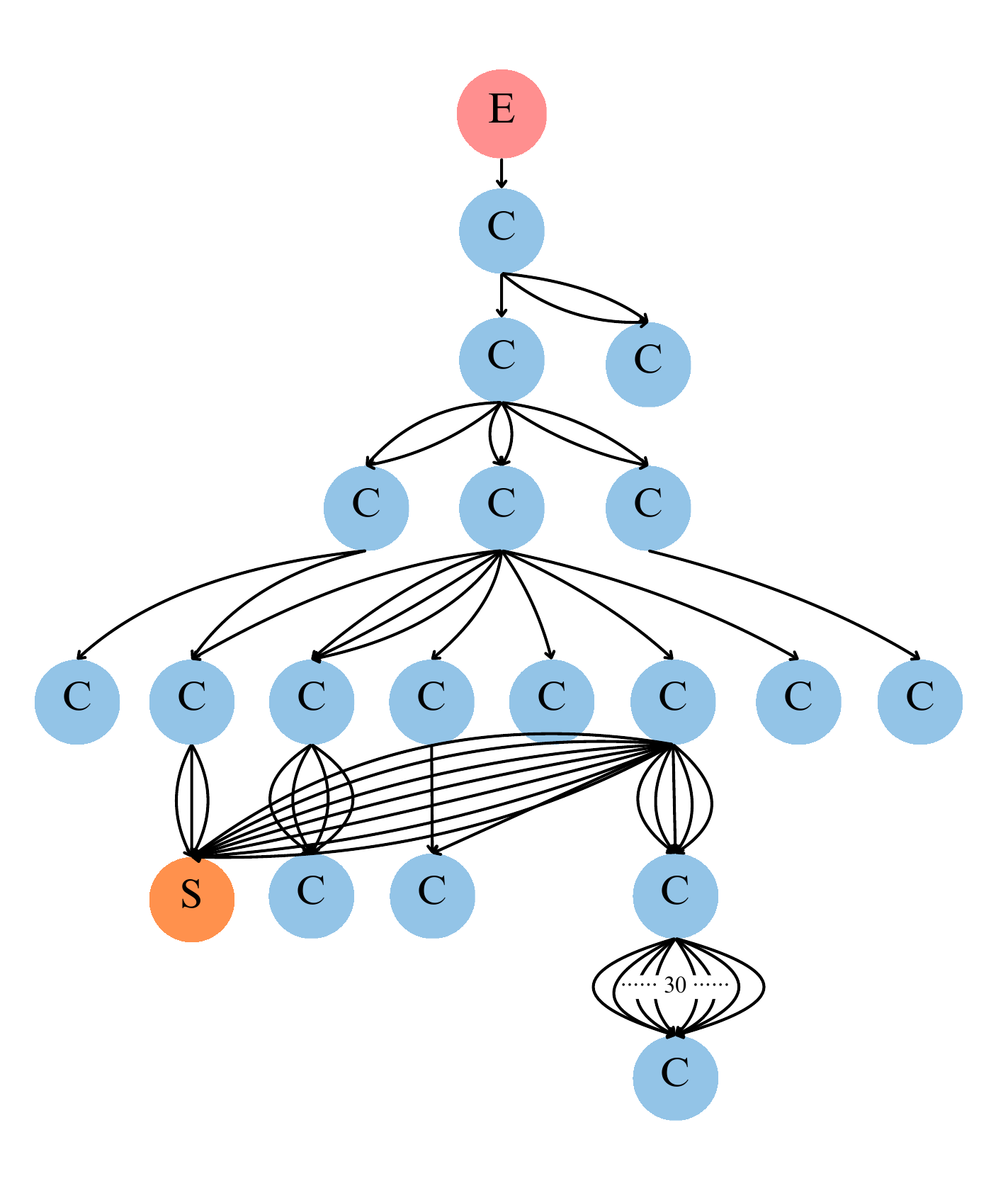}}

     \caption{Interaction Patterns between Smart Contracts Predominantly Observed in Rugpull-, Honeypot-, and Ponzi-Related Transactions.}
    \label{fig:scam_type_specific_pattern}
\end{figure}

Moreover, certain interaction patterns between smart contracts are predominantly observed in transactions involving specific scam tokens. In particular, 10,893 patterns are exclusively linked to rugpull tokens, 1,160 to Ponzi tokens, and 1,559 to honeypot tokens. Figure~\ref{fig:scam_type_specific_pattern} presents three representative patterns, which appear in 468,058, 2,202, and 1,424 transactions involving rugpull, Ponzi, and honeypot, respectively. These representative patterns are more frequently observed in risky transactions of their corresponding scam types, where they often exhibit relatively higher frequencies within each category.

\rqbox{
\textbf{Finding 5:} 
The ten most frequently observed interaction patterns between smart contracts in risky transactions can be classified into two categories: 1) those that are prevalent in both risky and non-risky transactions, and 2) those predominantly associated with risky transactions, characterized by isolated surges in transaction volumes across distinct short intervals from April 2018 to July 2024. The transaction volumes associated with all patterns demonstrate a consistent decline after early 2022.
}

\section{Discussion}\label{sec:discussion}
\subsection{Implications}
We reflect on our findings from the research questions, offering practical recommendations to mitigate risks in the NFT ecosystem. We also highlight the avenues of future research.

\noindent\textbf{Impact and Risks of EIP-1967 Proxy Contracts.} In RQ1, we observed that EIP-1967 proxy contracts represent three of the top ten semantic clusters of smart contracts deployed on Ethereum in the NFT ecosystem. Meanwhile, proxy contracts are among the top five most interactive smart contracts during NFT transactions, as observed in RQ2. 
Given the prevalence of EIP-1967 proxy contracts in the NFT ecosystem, their reliability and stability are critical to the overall security and operation of the NFT ecosystem. 
Malicious actors may exploit proxy contracts to conceal malicious logic, misleading blockchain explorers (e.g., Etherscan) and monitoring tools into mistakenly identifying them as legitimate logic contracts~\cite{community2023evasion}, particularly in the context of EIP-1967.
Specifically, malicious proxies can store the address of a legitimate logic contract in the designated slot defined by EIP-1967, while redirecting execution to a separate logic contract that contains malicious logic at runtime~\cite{zhang2025empirical}.
Future work could put efforts into developing a real-time monitoring system of logic contracts behind proxies during NFT transactions, enabling the early detection of anomalous or suspicious activities.

\noindent\textbf{Prevalence of Cloned EIP-1167 Proxy Contracts.} In RQ1, we observed that EIP-1167 contracts, acting as minimal proxies and are not upgradeable by design, account for 74.82\% of the deployed smart contracts in the NFT ecosystem, far exceeding the the 19.54\% share of non-upgradeable proxies reported on Ethereum~\cite{zhang2025empirical}.
The majority of these cloned EIP-1167 proxy contracts are configured to delegate calls to a smart contract known as \emph{XEN Torrent}~\cite{XENTorrent}. XEN Torrent facilitates bulk minting of XEN tokens through a mechanism involving Virtual Minting Units (VMUs)~\cite{xencrypto2022report}. Users can create VMUs by interacting with XEN Torrent, which are then used to claim XEN tokens.
The centralization around XEN Torrent could expose the NFT ecosystem to systemic risks if vulnerabilities in these contracts are exploited.

\noindent\textbf{Fraud Detection and Transaction Risk Assessment.} 
In RQ3, we observed that scam tokens are distributed across 45 out of 1,350 (3.3\%) bytecode clusters, with 38 of these clusters exhibiting high scam token ratios exceeding 60\%, indicating a semantic convergence of scam tokens in their bytecodes.
Such convergence could inform the development of cluster-based approaches for scam token detection in the NFT ecosystem.
Moreover, previous studies~\cite{wu2024tokenscout,zipzap,detectmaliciousaccount,CIKM24} have proposed graph-based models for fraud detection in the Ethereum ecosystem, where tokens are represented as nodes in the graphs.
Future work could explore the potential of using the distinct differences in the bytecodes between scam and non-scam tokens to enhance graph-based fraud detection models, particularly in the process of node embedding.

We also identified four interaction patterns--S2, S7, S9, and S10--between smart contracts in risky transactions, exhibiting risky transaction ratios existing 90\% (RQ3). 
These interaction patterns exhibit strong discriminatory power in differentiating between risky and non-risky transactions, and could therefore serve as behavioral signatures of fraudulent activities.
Future work can incorporate these patterns as high-level semantic features into risk assessment frameworks for transactions in the NFT ecosystem.

\subsection{Threats to Validity}
Our study exclusively focuses on the ERC721 token standard, which enables a more targeted investigation, but may limit the generalizability of our findings to other NFT token standards, such as ERC1155. In RQ1, the labeling of semantic clusters of smart contracts was independently conducted by two authors to reduce personal biases; however, the process may still involve a degree of subjectivity. To mitigate this threat, a blockchain expert was consulted to validate the labeling results, thereby enhancing the reliability of the annotations.
Another potential threat to validity arises from the lack of sufficient off-chain data, particularly in the analysis presented in RQ3. Risky transactions may involve complex, multi-step interactions that occur off-chain before assets are transferred to centralized exchanges, mixers, or cross-chain bridges. The inability to access comprehensive off-chain information limits our capacity to fully trace and analyze these transactions, potentially omitting factors that influence transaction patterns and their associations with illicit activities.

\section{Related Work}\label{sec:related}
Some researchers have conducted empirical studies to explore the NFT ecosystem from various perspectives, including security challenges~\cite{das2022understanding}, user activities on the prominent NFT marketplace~\cite{white2022characterizing}, and on-chain behaviors of NFTs~\cite{huang2024unveiling}. Specifically, Das et al.~\cite{das2022understanding} explored the security challenges in the NFT ecosystem with respect to the participating actors, including flaws in NFT marketplace design, threats from off-chain external entities, and trading malpractices by malicious users. 
White et al.~\cite{white2022characterizing} performed a longitudinal measurement study on the OpenSea NFT marketplace, focusing on the activities of buyers and sellers across various categories of NFT collections on OpenSea.
A more recent work~\cite{huang2024unveiling} conducted a large-scale measurement study of the NFT ecosystem, analyzing the on-chain behaviors of NFTs through the lens of events, participants, and marketplaces, as well as leveraging NFT trading data to investigate market manipulation, such as wash trading and arbitrage.
Our work primarily focuses on the semantics and interactions of smart contracts during NFT transactions, thereby distinguishing it from the aforementioned studies that address other facets of the NFT ecosystem.

Other researchers have focused on specific risks in the NFT ecosystem, including rug pulls~\cite{huang2023miracle}, wash trading~\cite{von2022nft,la2023game}, promotion scams~\cite{roy2024unveiling}, and phishing scams~\cite{yang2024stole}. 
Specifically, Huang et al.~\cite{huang2023miracle} characterized the symptoms of NFT rug pulls and proposed approaches for automatic detection and early warning of such scams.
Regarding NFT wash trading, von Wachter et al.~\cite{von2022nft} proposed an approach to identify suspicious wash trading behaviors in transactions on the OpenSea NFT marketplace. Later, La Morgia et al.~\cite{la2023game} performed a systematic analysis of NFT wash trading across six Ethereum-based NFT marketplaces, and measured the profitability of wash trading activities.
As for NFT promotion scams, Roy et al.~\cite{roy2024unveiling} conducted a longitudinal study on fraudulent NFT project promotions on Twitter, characterizing the associated Twitter accounts and the tactics used in these scams. They further proposed a machine learning model to identify fraudulent projects promoted on Twitter.
In terms of NFT phishing scams, Yang et al.~\cite{yang2024stole} explored the patterns and economic impact of phishing scams, as well as the NFT preference and post-theft behavior of scammers. They also proposed approaches for detecting NFT phishing accounts. 
In contrast to the aforementioned prior studies that focus on specific risks, our work investigates how scam token risks manifest through the semantics and interactions of smart contracts during transactions.

\section{conclusion and Future Work}\label{sec:conclusion}
In this work, we conducted a large-scale empirical study to explore the semantics and interactions of smart contracts in NFT transactions, using a curated dataset of nearly 100 million transactions across 20 million blocks on Ethereum.
We characterize the semantics of smart contracts in the NFT ecosystem, analyze the frequency and complexity of interactions between smart contracts during NFT transactions, and explore how the risks of scam tokens manifest with respect to the semantics and interactions of contracts.
Future work could put efforts into improving the security of proxy smart contracts through real-time monitoring of their corresponding logic contracts, the development of graph-based models for fraud detection by taking into account bytecode features, as well as the integration of smart contract interaction patterns into transaction risk assessment frameworks.

\section{Acknowledgement}
This research was supported by the National Science Foundation of China (No. 62472383), the Fundamental Research Funds for the Central Universities (No. 226-2025-00004), and the Open Research Fund of the State Key Laboratory of Blockchain and Data Security, Zhejiang University.

\bibliographystyle{IEEEtran}

\bibliography{nft}

@InProceedings{campello2013hdbscan,
author="Campello, Ricardo J. G. B.
and Moulavi, Davoud
and Sander, Joerg",
editor="Pei, Jian
and Tseng, Vincent S.
and Cao, Longbing
and Motoda, Hiroshi
and Xu, Guandong",
title="Density-Based Clustering Based on Hierarchical Density Estimates",
booktitle="Advances in Knowledge Discovery and Data Mining",
year="2013",
publisher="Springer Berlin Heidelberg",
address="Berlin, Heidelberg",
pages="160--172",
isbn="978-3-642-37456-2"
}

@misc{decrypt2021,
  title={Now postage stamps are getting the NFT treatment},
  author={{Decrypt}},
  year={2021},
  howpublished={\url{https://decrypt.co/61963/now-postage-stamps-are-getting-the-nft-treatment}},
}

@misc{blockchainappfactory,
  title={Real estate tokenization},
  author={{Blockchain App Factory}},
  howpublished={\url{https://www.blockchainappfactory.com/real-estate-tokenization}},
}

@misc{prnewswire2021,
  title={USPS certifies Casemail as first blockchain generated epostage},
  author={{PR Newswire}},
  year={2021},
  howpublished={\url{https://www.prnewswire.com/news-releases/usps-certifies-casemail-as-first-blockchain-generated-epostage-301267842.html}},
}

@misc{cointelegraph2021,
  title={You can now buy gold-backed NFTs with the mining carbon footprint offset},
  author={{CoinTelegraph}},
  year={2021},
  howpublished={\url{https://cointelegraph.com/news/you-can-now-buy-gold-backed-nfts-with-the-mining-carbon-footprint-offset}},
}

@article{wang2021,
  title={Non-fungible token (NFT): Overview, evaluation, opportunities and challenges},
  author={Qin Wang and Rujia Li and Qi Wang and Shiping Chen},
  journal={arXiv preprint arXiv:2105.07447},
  year={2021}
}

@misc{flipkick,
  title={Flipkick},
  howpublished={\url{https://www.flipkick.io}},
  note={Accessed: 2025-05-25},
  author={Flipkick}
}

@misc{coingecko2023annual,
  title        = {2023 Annual Crypto Industry Report},
  author       = {CoinGecko},
  year         = {2024},
  note         = {Accessed: 2025-05-25},
  url          = {https://assets.coingecko.com/reports/2023/CoinGecko-2023-Annual-Crypto-Industry-Report.pdf}
}

@inproceedings{das2022understanding,
  title={Understanding security issues in the NFT ecosystem},
  author={Das, Dipanjan and Bose, Priyanka and Ruaro, Nicola and Kruegel, Christopher and Vigna, Giovanni},
  booktitle={Proceedings of the 2022 ACM SIGSAC Conference on Computer and Communications Security},
  pages={667--681},
  year={2022}
}

@inproceedings{yang2023definition,
  title={Definition and detection of defects in NFT smart contracts},
  author={Yang, Shuo and Chen, Jiachi and Zheng, Zibin},
  booktitle={Proceedings of the 32nd ACM SIGSOFT International Symposium on Software Testing and Analysis},
  pages={373--384},
  year={2023}
}

@inproceedings{white2022characterizing,
  title={Characterizing the OpenSea NFT marketplace},
  author={White, Bryan and Mahanti, Aniket and Passi, Kalpdrum},
  booktitle={Companion Proceedings of the Web Conference 2022},
  pages={488--496},
  year={2022}
}

@article{wu2021defiranger,
  title={Defiranger: Detecting price manipulation attacks on defi applications},
  author={Wu, Siwei and Wang, Dabao and He, Jianting and Zhou, Yajin and Wu, Lei and Yuan, Xingliang and He, Qinming and Ren, Kui},
  journal={arXiv preprint arXiv:2104.15068},
  year={2021}
}

@misc{geth,
    note = {\url{https://geth.ethereum.org}},
    title = {Geth},
    year = {2023}
}

@article{kuhn2007semantic,
  title={Semantic clustering: Identifying topics in source code},
  author={Kuhn, Adrian and Ducasse, St{\'e}phane and G{\^\i}rba, Tudor},
  journal={Information and software technology},
  volume={49},
  number={3},
  pages={230--243},
  year={2007},
  publisher={Elsevier}
}

@inproceedings{liang2023needle,
  title={A Needle is an Outlier in a Haystack: Hunting Malicious PyPI Packages with Code Clustering},
  author={Liang, Wentao and Ling, Xiang and Wu, Jingzheng and Luo, Tianyue and Wu, Yanjun},
  booktitle={2023 38th IEEE/ACM International Conference on Automated Software Engineering (ASE)},
  pages={307--318},
  year={2023},
  organization={IEEE}
}

@inproceedings{shahapure2020cluster,
  title={Cluster quality analysis using silhouette score},
  author={Shahapure, Ketan Rajshekhar and Nicholas, Charles},
  booktitle={2020 IEEE 7th international conference on data science and advanced analytics (DSAA)},
  pages={747--748},
  year={2020},
  organization={IEEE}
}

@inproceedings{huang2024unveiling,
  title={Unveiling the paradox of nft prosperity},
  author={Huang, Jintao and Xia, Pengcheng and Li, Jiefeng and Ma, Kai and Tyson, Gareth and Luo, Xiapu and Wu, Lei and Zhou, Yajin and Cai, Wei and Wang, Haoyu},
  booktitle={Proceedings of the ACM Web Conference 2024},
  pages={167--177},
  year={2024}
}

@inproceedings{roy2024unveiling,
  title={Unveiling the risks of NFT promotion scams},
  author={Roy, Sayak Saha and Das, Dipanjan and Bose, Priyanka and Kruegel, Christopher and Vigna, Giovanni and Nilizadeh, Shirin},
  booktitle={Proceedings of the International AAAI Conference on Web and Social Media},
  volume={18},
  pages={1367--1380},
  year={2024}
}

@inproceedings{von2022nft,
  title={NFT wash trading: Quantifying suspicious behaviour in NFT markets},
  author={von Wachter, Victor and Jensen, Johannes Rude and Regner, Ferdinand and Ross, Omri},
  booktitle={International Conference on Financial Cryptography and Data Security},
  pages={299--311},
  year={2022},
  organization={Springer}
}

@article{yang2024stole,
  title={Who Stole My NFT? Investigating Web3 NFT Phishing Scams on Ethereum},
  author={Yang, Jingjing and Liu, Jieli and Lin, Dan and Wu, Jiajing and Huang, Baoying and Li, Quanzhong and Zheng, Zibin},
  journal={IEEE Transactions on Information Forensics and Security},
  year={2024},
  publisher={IEEE}
}

@article{huang2023miracle,
  title={Miracle or mirage? a measurement study of nft rug pulls},
  author={Huang, Jintao and He, Ningyu and Ma, Kai and Xiao, Jiang and Wang, Haoyu},
  journal={Proceedings of the ACM on Measurement and Analysis of Computing Systems},
  volume={7},
  number={3},
  pages={1--25},
  year={2023},
  publisher={ACM New York, NY, USA}
}

@inproceedings{la2023game,
  title={A game of NFTs: Characterizing NFT wash trading in the Ethereum blockchain},
  author={La Morgia, Massimo and Mei, Alessandro and Mongardini, Alberto Maria and Nemmi, Eugenio Nerio},
  booktitle={2023 IEEE 43rd International Conference on Distributed Computing Systems (ICDCS)},
  pages={13--24},
  year={2023},
  organization={IEEE}
}

@article{murray2018eip,
  title={EIP-1167: minimal proxy contract},
  author={Murray, Peter and Welch, Nate and Messerman, Joe},
  journal={ethereum improvement proposals},
  number={1167},
  year={2018}
}

@misc{erc721a,
  title        = {ERC721A},
  year         = {2021},
  howpublished = {\url{https://github.com/chiru-labs/ERC721A}},
}

@misc{uniswapv3,
    author={Adams, Hayden and Zinsmeister, Noah and Salem, Moody and Keefer, River and Robinson, Dan},
    title={Uniswap v3 Core},
    year={2021},
    howpublished={\url{https://app.uniswap.org/whitepaper-v3.pdf}},
    note= {Accessed: 2025-05-25}
}

@inproceedings{wu2024tokenscout,
  title={Tokenscout: Early detection of ethereum scam tokens via temporal graph learning},
  author={Wu, Cong and Chen, Jing and Zhao, Ziming and He, Kun and Xu, Guowen and Wu, Yueming and Wang, Haijun and Li, Hongwei and Liu, Yang and Xiang, Yang},
  booktitle={Proceedings of the 2024 on ACM SIGSAC Conference on Computer and Communications Security},
  pages={956--970},
  year={2024}
}

@inproceedings{zipzap,
author = {Hu, Sihao and Huang, Tiansheng and Chow, Ka-Ho and Wei, Wenqi and Wu, Yanzhao and Liu, Ling},
title = {ZipZap: Efficient Training of Language Models for Large-Scale Fraud Detection on Blockchain},
year = {2024},
isbn = {9798400701719},
publisher = {Association for Computing Machinery},
address = {New York, NY, USA},
url = {https://doi.org/10.1145/3589334.3645352},
doi = {10.1145/3589334.3645352},
booktitle = {Proceedings of the ACM Web Conference 2024},
pages = {2807–2816},
numpages = {10},
location = {Singapore, Singapore},
series = {WWW '24}
}

@inproceedings{detectmaliciousaccount,
author = {Li, Wenkai and Liu, Zhijie and Li, Xiaoqi and Nie, Sen},
title = {Detecting Malicious Accounts in Web3 through Transaction Graph},
year = {2024},
isbn = {9798400712487},
publisher = {Association for Computing Machinery},
address = {New York, NY, USA},
url = {https://doi.org/10.1145/3691620.3695344},
doi = {10.1145/3691620.3695344},
booktitle = {Proceedings of the 39th IEEE/ACM International Conference on Automated Software Engineering},
pages = {2482–2483},
numpages = {2},
location = {Sacramento, CA, USA},
series = {ASE '24}
}

@inproceedings{CIKM24,
author = {Ding, Zhihao and Shi, Jieming and Li, Qing and Cao, Jiannong},
title = {Effective Illicit Account Detection on Large Cryptocurrency MultiGraphs},
year = {2024},
isbn = {9798400704369},
publisher = {Association for Computing Machinery},
address = {New York, NY, USA},
url = {https://doi.org/10.1145/3627673.3679707},
doi = {10.1145/3627673.3679707},
booktitle = {Proceedings of the 33rd ACM International Conference on Information and Knowledge Management},
pages = {457–466},
numpages = {10},
location = {Boise, ID, USA},
series = {CIKM '24}
}

@misc{2024rugpull,
    title={Rugpull},
    year={2024},
    howpublished={\url{https://coinmarketcap.com/alexandria/glossary/rug-pull}}
}

@inproceedings{cernera2023token,
  title={Token spammers, rug pulls, and sniper bots: An analysis of the ecosystem of tokens in ethereum and in the binance smart chain ($\{$$\{$$\{$$\{$$\{$BNB$\}$$\}$$\}$$\}$$\}$)},
  author={Cernera, Federico and La Morgia, Massimo and Mei, Alessandro and Sassi, Francesco},
  booktitle={32nd USENIX security symposium (USENIX security 23)},
  pages={3349--3366},
  year={2023}
}

@inproceedings{torres2019art,
  title={The art of the scam: Demystifying honeypots in ethereum smart contracts},
  author={Torres, Christof Ferreira and Steichen, Mathis and others},
  booktitle={28th USENIX Security Symposium (USENIX Security 19)},
  pages={1591--1607},
  year={2019}
}

@misc{2024ponzi,
    title={Ethereum’s top gas guzzlers are ponzi schemes},
    year={2024},
    howpublished={\url{https://cryptonews.net/ news/ethereum/384739/}}
}

@misc{community2023evasion,
  author = {Fonta Community},
  title = {Evasion techniques: Report on the continuous monitoring},
  year = {2023},
  howpublished = {\url{https://github.com/apehex/web3-evasion-techniques/blob/main/report/forta.pdf}},
  note = {Accessed: 2025-05-25}
}

@inproceedings{zhang2025empirical,
  author = {Zhang, Mengya and Shukla, Preksha and Zhang, Wuqi and Zhang, Zhuo and Agrawal, Pranav and Lin, Zhiqiang and Zhang, Xiangyu and Zhang, Xiaokuan},
  title = {An Empirical Study of Proxy Smart Contracts at the Ethereum Ecosystem Scale},
  booktitle = {2025 IEEE/ACM 47th International Conference on Software Engineering (ICSE)},
  year = {2025},
  pages = {620--620},
  doi = {10.1109/ICSE55347.2025.00083},
}

@misc{xencrypto2022report,
  author = {{XEN Crypto}},
  title = {XEN Crypto: Whitepaper},
  year = {2022},
  url = {https://www.xencrypto.io/wp-content/uploads/2022/12/REP-final-20221227T163457Z.pdf},
  note = {Accessed: 2025-05-25}
}

@article{zheng2020xblock,
  title={Xblock-eth: Extracting and exploring blockchain data from ethereum},
  author={Zheng, Peilin and Zheng, Zibin and Wu, Jiajing and Dai, Hong-Ning},
  journal={IEEE Open Journal of the Computer Society},
  volume={1},
  pages={95--106},
  year={2020},
  publisher={IEEE}
}

@misc{business2024losses,
  author = {{The Business Times}},
  title = {Losses from crypto scams grew 45\% in 2023, FBI says},
  year = {2024},
  howpublished = {\url{https://www.businesstimes.com.sg/companies-markets/banking-finance/losses-crypto-scams-grew-45-2023-fbi-says}},
  note = {Accessed: 2025-05-25}
}

@misc{erc721oppenzeppelin,
  author = {OpenZeppelin},
  title = {ERC721},
  year = {2024},
  howpublished = {\url{https://docs.openzeppelin.com/contracts/4.x/api/token/erc721}},
  note = {Accessed: 2025-05-25}
}

@misc{wyvern,
  author = {{Wyvern Protocol}},
  title = {Wyvern Protocol},
  year = {2019},
  howpublished = {\url{https://wyvernprotocol.com/docs/protocol-components}},
  note = {Accessed: 2025-05-25}
}

@misc{seaport,
  author = {OpenSea},
  title = {Seaport},
  year = {2024},
  howpublished = {\url{https://docs.opensea.io/docs/seaport}},
  note = {Accessed: 2025-05-25}
}

@misc{safe,
  author = {Safe},
  title = {What is Safe?},
  year = {2025},
  howpublished = {\url{https://docs.safe.global/home/what-is-safe}},
  note = {Accessed: 2025-05-25}
}

@misc{manifold,
  author = {Manifold},
  title = {Manifold},
  year = {2023},
  howpublished = {\url{https://docs.manifold.xyz/manifold-for-developers}},
  note = {Accessed: 2025-05-25}
}

@misc{alchemymanifold,
  author = {Alchemy},
  title = {Manifold},
  year = {2024},
  howpublished = {\url{https://www.alchemy.com/dapps/manifold}},
  note = {Accessed: 2025-05-25}
}

@misc{jpegmining,
  author = {{Buterin Cards}},
  title = {Buterin Cards},
  year = {2024},
  howpublished = {\url{https://www.buterin.cards}},
  note = {Accessed: 2025-05-25}
}

@misc{etherscan,
  author       = {Etherscan},
  title        = {Etherscan},
  howpublished = {\url{https://etherscan.io}},
  note         = {Accessed: 2025-05-25}
}

@misc{bge,
  author       = {BAAI},
  title        = {bge-large-en-v1.5},
  howpublished = {\url{https://huggingface.co/BAAI/bge-large-en-v1.5}},
  note         = {Accessed: 2025-05-25}
}

@misc{hdbscan,
  title        = {hdbsan},
  howpublished = {\url{https://pypi.org/project/hdbscan/}},
  note         = {Accessed: 2025-05-25}
}

@misc{tx_p10,
  howpublished = {\url{https://etherscan.io/tx/0xbbdf118861b0539bd9f19958dd1bd437c39b10c- 2bc06c96307f796da2fb2f5e2}},
  note         = {Accessed: 2025-05-25}
}

@misc{leftoverz,
    title = {leftoverz},
    howpublished = {\url{https://etherscan.io/address/0xb8f9bfC712E77F9DfFE22EA8f9ADaE8d4314d0D2#code}},
  note         = {Accessed: 2025-05-25}
}

@misc{CryptoToon,
    title = {CryptoToon},
    howpublished = {\url{https://etherscan.io/address/0xBF8a84DE5DcC0bd5792026BFDeBFc75d9675A363#code}},
  note         = {Accessed: 2025-05-25}
}

@misc{codex,
    title = {CodexRecordProxy},
    howpublished = {\url{https://etherscan.io/address/0x8853B05833029e3Cf8d3Cbb592f9784FA43d2a79#code}},
  note         = {Accessed: 2025-05-25}
}

@misc{MLBNFT,
    title = {MLBNFT},
    howpublished = {\url{https://etherscan.io/address/0x8c9b261Faef3b3C2e64ab5E58e- 04615F8c788099#code}},
    note = {Accessed: 2025-05-25}
}

@misc{MarbleNFT,
    title = {MarbleNFT},
    howpublished = {\url{https://etherscan.io/address/0x1d963688FE2209A98dB35C67A041524822Cf04ff#code}},
    note = {Accessed: 2025-05-25}
}

@misc{CryptoFlower,
    name = {CryptoFlower},
    howpublished = {\url{https://etherscan.io/address/0x65FCFB6870C744Ec181e4F64a7F41A0C-  fd76B845#code}},
    note = {Accessed: 2025-05-25}
}

@misc{XENTorrent,
    name = {XENTorrent},
    howpublished = {\url{https://etherscan.io/address/0x0a252663dbcc0b073063d6420a40319e438cfa59#code}},
    note = {Accessed: 2025-05-25}
}

@misc{replication_package,
  author       = {Anonymous},
  title        = {Replication Package for `Why Is My Transaction Risky? Understanding Smart Contract Semantics and Interactions in the NFT Ecosystem'},
  year         = {2025},
  publisher    = {Zenodo},
  howpublished = {\url{https://doi.org/10.5281/zenodo.15550314}}
}

@inproceedings{
zhang2025harnessing,
title={Harnessing Diversity for Important Data Selection in Pretraining Large Language Models},
author={Chi Zhang and Huaping Zhong and Kuan Zhang and Chengliang Chai and Rui Wang and Xinlin Zhuang and Tianyi Bai and Qiu Jiantao and Lei Cao and Ju Fan and Ye Yuan and Guoren Wang and Conghui He},
booktitle={The Thirteenth International Conference on Learning Representations},
year={2025},
url={https://openreview.net/forum?id=bMC1t7eLRc}
}

@inproceedings{liang-etal-2024-controlled,
    title = "Controlled Text Generation for Large Language Model with Dynamic Attribute Graphs",
    author = "Liang, Xun  and
      Wang, Hanyu  and
      Song, Shichao  and
      Hu, Mengting  and
      Wang, Xunzhi  and
      Li, Zhiyu  and
      Xiong, Feiyu  and
      Tang, Bo",
    editor = "Ku, Lun-Wei  and
      Martins, Andre  and
      Srikumar, Vivek",
    booktitle = "Findings of the Association for Computational Linguistics: ACL 2024",
    month = aug,
    year = "2024",
    address = "Bangkok, Thailand",
    publisher = "Association for Computational Linguistics",
    url = "https://aclanthology.org/2024.findings-acl.345/",
    doi = "10.18653/v1/2024.findings-acl.345",
    pages = "5797--5814"
}

@inproceedings{ma2025surviving,
  title={Surviving in Dark Forest: Towards Evading the Attacks from Front-Running Bots in Application Layer},
  author={Ma, Zuchao and Jiang, Muhui and Luo, Feng and Luo, Xiapu and Zhou, Yajin},
  booktitle={34rd USENIX Security Symposium (USENIX Security 25)},
  year={2025}
}

@article{zheng2025gnncontext,
  title={GNNCONTEXT: GNN-based Code Context Prediction for Programming Tasks},
  author={Zheng, Xiaoye and Wan, Zhiyuan and Liu, Shun and Yang, Kaiwen and Lo, David and Yang, Xiaohu},
  journal={IEEE Transactions on Software Engineering},
  year={2025},
  publisher={IEEE}
}

@article{wang2025clone,
  title={Clone Detection for Smart Contracts: How Far Are We?},
  author={Wang, Zuobin and Wan, Zhiyuan and Chen, Yujing and Zhang, Yun and Lo, David and Xie, Difan and Yang, Xiaohu},
  journal={Proceedings of the ACM on Software Engineering},
  volume={2},
  number={FSE},
  pages={1249--1269},
  year={2025},
  publisher={ACM New York, NY, USA}
}

\end{document}